\documentstyle[prl,aps,graphics,floats]{revtex}

\newcommand{\DegC}{\char'27\kern-.3em\hbox{C}}

\begin{document}

\draft

\title{Testing Lorentz  and CPT symmetry 
with hydrogen masers}

\author{M.A. Humphrey, D.F. Phillips, E.M. Mattison, R.F.C. Vessot, 
R.E. Stoner and R.L. Walsworth}
\address{Harvard-Smithsonian Center for Astrophysics, Cambridge, MA 02138}

\date{\today}

\maketitle

\begin{abstract}

We present details from a recent test of Lorentz and CPT symmetry 
using hydrogen masers \cite{lli.Hexp}.   We have placed a new limit on 
Lorentz and CPT violation of the proton in terms of a recent standard model 
extension by placing a bound on 
sidereal variation of the $F=1$, $\Delta m_{F}=\pm1$ Zeeman frequency 
in hydrogen.  Here, the theoretical standard model extension is reviewed.  
The operating principles of the maser 
and the double resonance technique used to measure the Zeeman frequency are discussed.   
The characterization of systematic effects is described, and the method of data 
analysis is presented.  We compare our result to other recent 
experiments, and discuss potential steps to improve our measurement.

\end{abstract}



\section{Introduction}
\label{sec.intro}

A theoretical framework has recently been developed that 
incorporates Lorentz and CPT symmetry violation into the standard 
model and quantifies their effects.
\cite{Kost.gen1,Kost.gen2,Kost.gen3,Kost.string1,Kost.string2,Kost.string3,Kost.string4,Kost.part1,Kost.part2,Kost.part3,Kost.part4,Kost.clock1,Kost.clock2}.  
One branch of this framework emphasizes low energy, experimental searches 
for symmetry violating effects in atomic energy levels \cite{Kost.clock1,Kost.clock2}.
In particular, Lorentz and CPT violation in hydrogen has been examined and 
sidereal variations in the $F=1$, $\Delta m_{F}=\pm1$ Zeeman frequency 
have been quantified \cite{lli.Hthe}.  Motivated by this work, we 
have conducted a search for sidereal variation in the hydrogen 
Zeeman frequency, and have placed a new clean bound 
of 10$^{-27}$ GeV on Lorentz and CPT violation of the proton 
\cite{lli.Hexp}.

Here we provide additional details of the theoretical framework, experiment and analysis.  
In Sec. \ref{sec.LLICPT} we discuss the standard model extension.  
In Sec. \ref{sec.maser} we describe 
the basic concepts of hydrogen maser operation and our Zeeman 
frequency measurement technique.  In Sec. \ref{sec.experiment} we describe the 
procedure used to collect data and extract a sidereal bound 
on the Zeeman frequency.   In Sec. \ref{sec.error} we 
describe efforts to reduce 
and characterize systematic effects. Finally, in Sec. 
\ref{sec.discussion} we compare our result to other clock-comparison 
tests of Lorentz and CPT symmetry, and discuss potential means of 
improving our measurement.

\section{Lorentz and CPT symmetry violation in the standard model}
\label{sec.LLICPT}

Experimental investigations of Lorentz symmetry provide important 
tests of the standard model of particle physics and general 
relativity.  While the standard model successfully describes particle 
phenomenology, it is believed to be the low energy limit of a 
fundamental theory that incorporates gravity.  This underlying theory 
may be Lorentz invariant, yet contain spontaneous 
symmetry-breaking that could result in small violations of Lorentz 
invariance and CPT at the level of the standard model.

A theoretical framework has been developed to 
describe Lorentz and CPT violation at the level of the standard model 
by Kosteleck\'{y} and coworkers 
\cite{Kost.gen1,Kost.gen2,Kost.gen3,Kost.string1,Kost.string2,Kost.string3,Kost.string4,Kost.part1,Kost.part2,Kost.part3,Kost.part4,Kost.clock1,Kost.clock2}.  
This standard-model extension is quite general:  it emerges as the 
low-energy limit of any underlying theory that generates the standard 
model and contains spontaneous Lorentz symmetry violation \cite{Kost.gen1,Kost.gen2,Kost.gen3}.
For example, such characteristics might emerge from string theory 
\cite{Kost.string1,Kost.string2,Kost.string3,Kost.string4}.  
A key feature of the standard model extension is that it is formulated 
at the level of the known elementary particles, and thus enables 
quantitative comparison of a wide array of searches for Lorentz and CPT 
violation \cite{Kost.part1,Kost.part2,Kost.part3,Kost.part4}. 

``Clock comparison experiments'' are searches for temporal variations
in atomic energy levels.  
According to the standard model extension considered here, 
Lorentz and CPT violation may produce shifts in certain atomic levels,
whose magnitude depends
on the orientation of the atom's quantization axis relative to a fixed 
inertial frame  \cite{Kost.clock1,Kost.clock2}.  Certain atomic 
transition frequencies, therefore, may exhibit sinusoidal variation as 
the earth rotates on its axis.  New limits can be placed on 
Lorentz and CPT violation by bounding sidereal variation of these atomic 
transition frequencies.

Specifically, the description of Lorentz and CPT violation is included 
in the relativistic Lagrange density of the constituent particles of the atom.  
For example, the modified electron Lagrangian becomes \cite{Kost.clock1}
\begin{equation}
    \label{eqn.lagrangian}
    {\mathcal{L}} = \frac{1}{2} i \bar{\psi} \Gamma_{\nu} \partial^{\nu}
    \psi  - \bar{\psi} 
    M \psi + {\mathcal L}_{int}^{QED}
\end{equation}
where 
\begin{equation}
    \label{eqn.gamma}
    \Gamma_{\nu} = \gamma_{\nu} + \left( c_{\mu \nu} \gamma^{\mu} + d_{\mu 
         \nu} \gamma_{5} \gamma^{\mu}  +  e_{\nu} + i f_{\nu} \gamma_{5} 
	 + \frac{1}{2}g_{\lambda \mu \nu} \sigma^{\lambda \mu} \right)
\end{equation}
and
\begin{equation}
    \label{eqn.m}
    M = m + \left( a_{\mu} \gamma^{\mu} + b_{\mu} \gamma_{5} 
    \gamma^{\mu} + \frac{1}{2} H_{\mu \nu} \sigma^{\mu \nu} \right).
\end{equation}
The parameters $a_{\mu}$, $b_{\mu}$, $c_{\mu \nu}$, $d_{\mu \nu}$, 
$e_{\nu}$, $f_{\nu}$, $g_{\lambda 
\mu \nu}$ and $H_{\mu \nu}$ represent possible vacuum expectation values of 
Lorentz tensors generated through spontaneous Lorentz symmetry 
breaking in an underlying theory.  These are absent in the standard 
model.  The parameters $a_{\mu}$, $b_{\mu}$, $e_{\nu}$, $f_{\nu}$ and $g_{\lambda 
\mu \nu}$ represent coupling strengths for terms that violate both
CPT and Lorentz symmetry, while $c_{\mu \nu}$, $d_{\mu \nu}$, 
and $H_{\mu \nu}$ violate 
Lorentz symmetry only.  An analogous expression exists for the 
modified proton and neutron Lagrangians (a superscript will be appended to 
differentiate between the sets of parameters).  
The standard model extension treats only the free particle properties 
of the constituent particles, estimating that all interaction effects 
will be of higher order \cite{Kost.clock1}.  As a result, the interaction 
term ${\mathcal L}_{int}^{QED}$ 
is unchanged from the conventional, Lorentz invariant, QED interaction term. 

Within this phenomenological framework, the values of these parameters 
are not calculable; instead, values must be determined 
experimentally.  The general nature of this theory ensures that 
different experimental searches may place bounds on different 
combinations of Lorentz and CPT violating terms, while
direct comparisons between these experiments are possible (see Table 
\ref{tab.compare} and Ref. \cite{Kost.clock1}).  

The leading-order Lorentz and CPT violating energy level shifts for a 
given atom are obtained by summing 
over the individual free particle shifts of the atomic constituents.
From the symmetry violating correction to the relativistic Lagrangian, 
a non-relativistic correction Hamiltonian $\delta h$ is found using standard 
field theory techniques \cite{Kost.clock1}.  Assuming Lorentz and CPT violating effects to
be small, the energy level shifts are calculated perturbatively by taking 
the expectation value of the correction Hamiltonian with respect to the 
unperturbed atomic states, leading to a shift in 
an atomic ($F,m_{F}$) sublevel given by \cite{Kost.clock1}
\begin{equation}
    \label{eqn.level.shift}
    \Delta E_{F,m_{F}} = \langle F, m_{F} | n_{e} \delta h_{e} + 
    n_{p} \delta h_{p} + n_{n} \delta h_{n} | F, m_{F} \rangle.
\end{equation}
Here $n_{w}$ is the number of each type of particle and $\delta 
h_{w}$ is the corresponding correction Hamiltonian.
Note that for most atoms, the interpretation of energy
level shifts in terms of this standard model extension is
reliant on the particular model used to describe the 
atomic nucleus (e.g., the Schmidt model).  One key advantage of a 
study in hydrogen is the simplicity of the nuclear structure (a single 
proton), with its results uncompromised by any nuclear model uncertainty.
 
Among the most recent clock comparison experiments 
are Penning trap tests by Dehmelt and co-workers with the electron and 
positron \cite{dehmelt1,dehmelt2} which place a limit on electron 
Lorentz and CPT violation at 10$^{-25}$ GeV.  A recent re-analysis by Adelberger, Gundlach, 
Heckel, and 
co-workers of existing data from the ``E\"{o}t-Wash II'' spin-polarized 
torsion pendulum \cite{adelberger1,adelberger2} has improved this to 
a level of 10$^{-29}$ GeV \cite{adelberger3}, the most
stringent bound to date on Lorentz and CPT violation of the 
electron.  A new limit on neutron Lorentz and 
CPT violation has been placed at 10$^{-31}$ GeV by Bear et al. \cite{bear} using a dual 
species noble gas maser and comparing Zeeman frequencies of $^{129}$Xe and 
$^{3}$He.  The current limit 
on Lorentz and CPT violation of the proton is 10$^{-27}$ GeV, as derived from an 
experiment by Lamoreaux and Hunter \cite{hunter.HgCs.lli} which compared Zeeman 
frequencies of $^{199}$Hg and $^{133}$Cs.

\begin{figure}
\begin{center}
\includegraphics{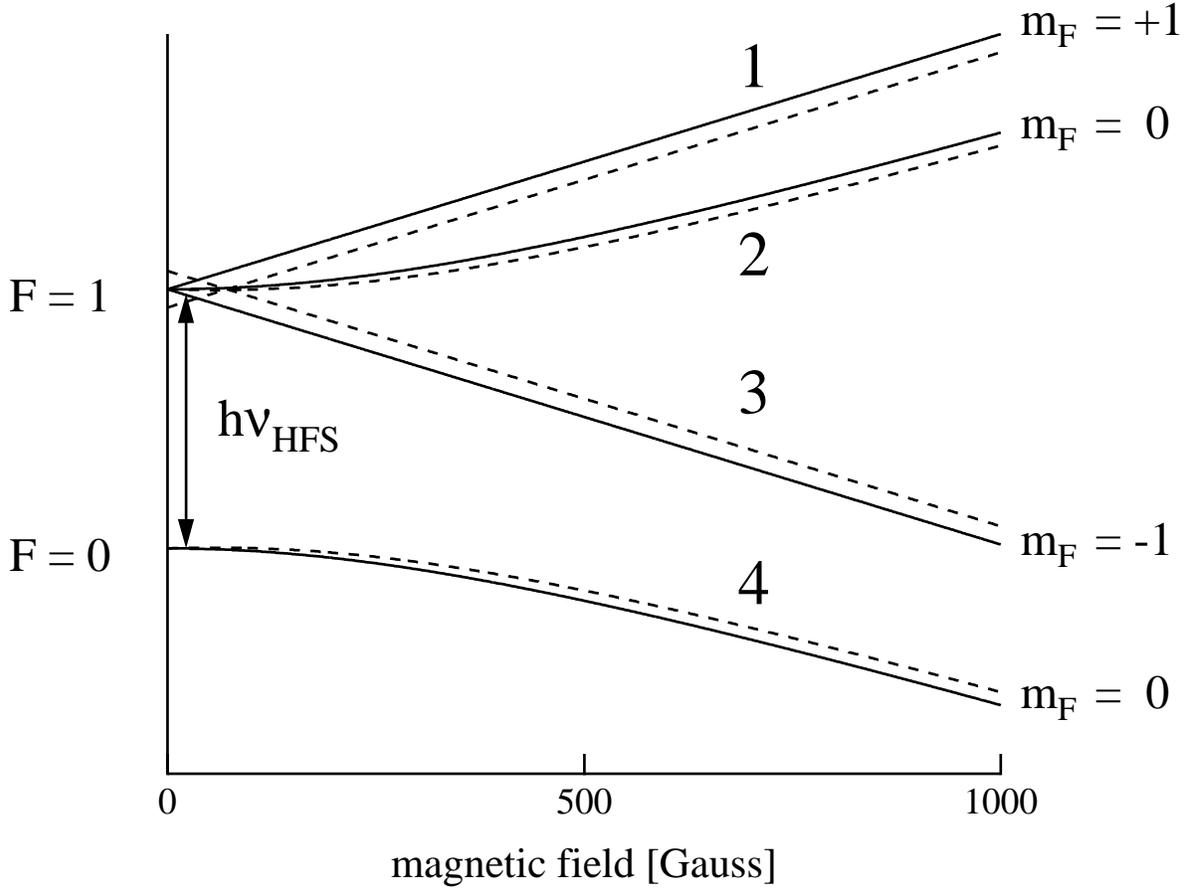}
\caption{Hydrogen hyperfine structure.  
The full curves are the unperturbed hyperfine levels, while the dashed curves 
illustrate the shifts due to Lorentz and CPT violating effects with 
the exaggerated values of $|b_{3}^{e} - 
    d_{30}^{e}m_{e}-H_{12}^{e}|$ = 90 MHz and $|b_{3}^{p} - 
    d_{30}^{p}m_{p}-H_{12}^{p}|$ = 10 MHz.  This work reports a 
bound of less than 1 mHz for these terms.  A hydrogen maser 
oscillates on the first-order magnetic field-independent $|2\rangle 
\leftrightarrow |4\rangle$ 
hyperfine transition near 1420 MHz.  The maser typically
operates with a static field less than 1 mG.  For these low field 
strengths, the two $F = 1$, $\Delta m_{F} = \pm 1$ Zeeman frequencies are 
nearly degenerate, and  $\nu_{12} \approx \nu_{23} \approx$ 1 kHz.}
\label{fig.levels}
\end{center}
\end{figure}

Figure \ref{fig.levels} shows the Lorentz and CPT violating 
corrections to the energy levels of the ground state of hydrogen \cite{lli.Hthe}.
The shift in the $F=1$, $\Delta m_{F} = \pm 1$ Zeeman frequency is \cite{gnote}:
\begin{equation}
    |\Delta \nu_{Z}| = \frac{1}{h} | (b_{3}^{e} - 
    d_{30}^{e}m_{e}-H_{12}^{e}) + (b_{3}^{p} - 
    d_{30}^{p}m_{p}-H_{12}^{p})|.
\label{eqn.Kost}
\end{equation}
The subscripts denote the projection of the tensor couplings 
onto the laboratory frame.  Therefore, as the earth rotates relative 
to a fixed inertial frame, the Zeeman frequency $\nu_{Z}$ will 
exhibit a sidereal variation.  
We have recently published the result of a search for this variation 
of the $F=1$, $\Delta m_{F} = \pm 1$ Zeeman frequency in hydrogen using 
hydrogen masers \cite{lli.Hexp}.  This search has placed a new, clean bound on 
Lorentz and CPT violation of the proton at a level of 10$^{-27}$ GeV.  

\section{Hydrogen maser concepts}
\label{sec.maser}  

The electronic ground state in hydrogen is split 
into four levels by the hyperfine interaction, labeled (following the 
notation of Andresen \cite{andresen}) $|1\rangle$ 
to $|4\rangle$ in order of decreasing energy (Fig.~\ref{fig.levels}).  The energies of  
atoms in $|1\rangle$ and $|2\rangle$ decrease as the magnetic 
field decreases; these are therefore low-field seeking states.  
Conversely, $|3\rangle$ and 
$|4\rangle$ are high-field seeking states.  In low fields, $|2\rangle$ and $|4\rangle$
are only dependent on magnetic field in second order.  The maser oscillates on the 
$|2\rangle \leftrightarrow |4\rangle$ transition (field-independent to first-order).  
This transition frequency, as a function of static field, is 
given by $\nu_{24} = \nu_{hfs} + 2750 B^{2}$ ($\nu$ in Hz with $B$ 
in Gauss, with $\nu_{hfs} \approx$ 1420.405751 MHz the zero-field hyperfine frequency).  
Hydrogen masers typically operate with low static fields (less than 1 mG), where 
$\nu_{24}$ is shifted from $\nu_{hfs}$
by about 3 mHz, or 2 parts in 10$^{12}$.  The two $F=1$, 
$\Delta m_{F} = \pm1$ Zeeman frequencies are 
given by $\nu_{12} = 1.4 \times 10^{6} B - 1375 
B^{2}$ and $\nu_{23} = 1.4 \times 10^{6} B + 1375 
B^{2}$.  At $B$ = 1 mG these are nearly degenerate, with $\nu_{12} - \nu_{23} 
\approx$ 3 mHz, much less than the Zeeman linewidth of approximately 1 Hz. 

\subsection{Maser operation}
\label{subsec.basics}

\begin{figure}
\begin{center}
\includegraphics{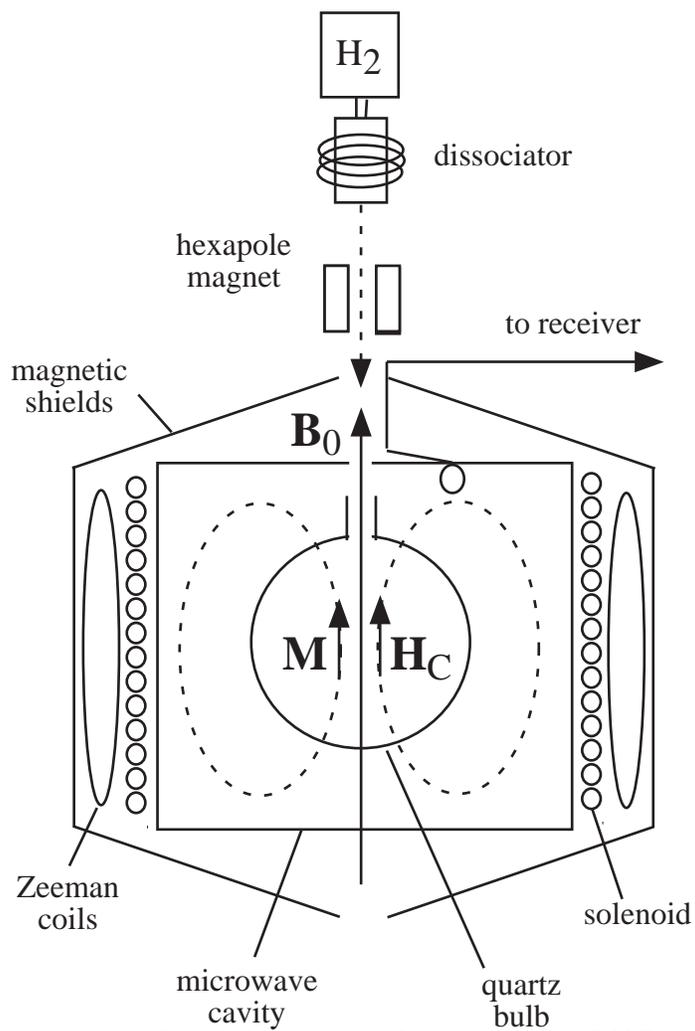}
\caption{Hydrogen maser schematic.  The solenoid generates a weak 
static magnetic field ${\mathbf B}_{0}$ which defines a quantization axis inside 
the maser bulb.  The microwave cavity field ${\mathbf H}_{C}$ (dashed 
field lines) and the coherent magnetization ${\mathbf M}$ of the atomic ensemble 
form the coupled actively oscillating system.}
\label{fig.schematic}
\end{center}
\end{figure}

In a hydrogen maser \cite{VanAud,Hmas1,Hmas2}, molecular hydrogen is
dissociated in an rf discharge and a beam of hydrogen atoms is
formed, as shown in Fig.~\ref{fig.schematic}.  A hexapole state
selecting magnet focuses the low-field-seeking
hyperfine states $|1\rangle$ and $|2\rangle$ into a quartz maser bulb
at about $10^{12}$ atoms/sec.  Inside the bulb (volume $\sim
10^{3}$ cm$^{3}$), the atoms travel ballistically for about 1 second
before escaping, making $\sim 10^{4}$ collisions with the bulb
wall.  A Teflon coating reduces the atom-wall interaction and thus
inhibits decoherence of the masing atomic ensemble by wall collisions.
The maser bulb is centered inside a cylindrical TE$_{011}$ microwave
cavity resonant with the 1420 MHz hyperfine transition.  The microwave
field stimulates a small, coherent magnetization in the atomic
ensemble, and this magnetization acts as a source to stimulate the
microwave field.  With sufficiently high atomic flux and low cavity
losses, this feedback induces active maser oscillation.  The maser
signal is inductively coupled out of the microwave cavity and
amplified with an external receiver.  Surrounding the cavity, a
solenoid produces the weak static magnetic field ($\approx$ 1 mG) that
establishes the quantization axis inside the maser bulb and sets the
Zeeman frequency ($\approx$ 1 kHz).  A pair of Helmholtz coils
produces the oscillating transverse magnetic field that drives the
$F=1$, $\Delta m_{F} = \pm1$ Zeeman transitions.  The cavity,
solenoid, and Zeeman coils are all enclosed within several
layers of high permeability magnetic shielding.

A well engineered hydrogen maser can have fractional stabilities 
approaching $10^{-15}$ over intervals of hours.   
This stability is enabled by a long atom-field interaction time (1 s), a 
low atom-wall interaction (due to the low atomic polarizability of H 
and the wall's Teflon coating), reduced Doppler effects (the atoms 
are confined to a region of uniform microwave field phase,
effectively averaging their velocity to zero over the interaction time with 
the field), and multiple layers of thermal control of the cavity 
(stabilizing cavity pulling shifts).  

\subsection{Maser characterization}
\label{subsec.numbers}

Among the quantities used to characterize a hydrogen 
maser, those most relevant to this experiment are the atomic 
line-Q $Q_{l}$, the population decay rate $\gamma_{1}$,
the hyperfine decoherence rate $\gamma_{2}$, 
the atomic flow rate into and out of the bulb $\gamma_{b}$, and the maser Rabi 
frequency $|X_{24}|$.  We describe here a comprehensive set of measurements to 
characterize hydrogen maser P-8.  The results discussed here are summarized 
in Table \ref{tab.char}.  Our Lorentz and CPT symmetry test data were taken 
with a similar but newer hydrogen maser, P-28 \cite{P8/P28}.  A few of the 
maser characterization parameters for P-28, while not directly measured, have 
been inferred using fitting parameters from the double resonance method used 
to measure the $F=1$, $\Delta m_{F}=\pm1$ 
Zeeman frequency, described in Sec. \ref{subsec.single}.  These values 
are included in Table \ref{tab.char} in italics. 

To determine these parameters of an operating H maser, the cavity volume $V_{C}$, bulb 
volume $V_{b}$, cavity quality factor $Q_{C}$, filling factor $\eta$, 
and output coupling coefficient $\beta$ must be known.  
For both masers, $V_{C}$ = 1.4 $\times$ 10$^{-2}$ 
m$^{3}$, $V_{b}$ = 2.9 $\times$ 10$^{-3}$ m$^{3}$, $Q_{C} \approx$ 
40,000, and $\beta$ = 0.23 \cite{ed.ieee}.  The filling factor, defined as 
\cite{Hmas1}
\begin{equation}
    \label{eqn.fill}
    \eta = \frac{\langle H_{z} \rangle _{bulb}^{2}}{\langle H^{2} \rangle _{cavity}},
\end{equation}
quantifies the ratio of average magnetic field energy inside the bulb to 
the average magnetic field energy in the cavity.  This has a value of $\eta$ = 
2.14 for masers P-8 and P-28 \cite{ed.ieee}.

\begin{table}
\begin{center}
\begin{tabular}{l c c c}
    parameter & symbol & P-8 & P-28 \\ 
    \hline
    cavity volume & $V_{C}$ & 1.4 $\times$ 10$^{-2}$ m$^{3}$ & 1.4 $\times$ 10$^{-2}$ m$^{3}$ \\ 
    bulb volume & $V_{b}$ & 2.9 $\times$ 10$^{-3}$ m$^{3}$ & 2.9 $\times$ 10$^{-3}$ m$^{3}$ \\
    cavity-Q & $Q_{C}$ & 39,346 & \\
    filling factor & $\eta$ & 2.14 & 2.14 \\
    line-Q & $Q_{l}$ & 1.6 $\times$ 10$^{9}$ & 1.6 $\times$ 10$^{9}$ \\
    maser quality parameter & $q$ & 0.100 & \\
    maser relaxation rate & $\gamma_{t}$ & 1.83 rad/s & \\
    bulb escape rate & $\gamma_{b}$ & 0.86 rad/s & 0.86 rad/s \\
    population decay rate & $\gamma_{1}$ & 4.04 rad/s & \emph{2.88 rad/s} \\
    maser decoherence rate & $\gamma_{2}$ & 2.77 rad/s & 2.8 rad/s \\
    spin-exchange decay rate & $\gamma_{se}$ & 1.06 rad/s & \\
    radiated power & $P$ & 600 fW & \\
    threshold power & $P_{c}$ & 250 fW \\
    output coupling & $\beta$ & 0.23 & 0.23 \\
    output power & $P_{o}$ & 112 fW & $\approx$ 100 fW \\
    total flux & $I_{tot}$ & 15.0 $\times$ 10$^{12}$ atoms/s & \\
    flux of $|2\rangle$ atoms & $I$ & 3.13 $\times$ 10$^{12}$ atoms/s & \\
    threshold flux & $I_{th}$ & 0.54 $\times$ 10$^{12}$ atoms/s & \\
    atomic density & $n$ & 2.8 $\times$ 10$^{15}$ atoms/m$^{3}$ & \\
    maser Rabi frequency & $|X_{24}|$ & 2.77 rad/s & \emph{2.14 rad/s} \\
\end{tabular}   
\caption{Maser characterization parameters.  The italicized values for P-28
were inferred from double resonance fit parameters as described in Sec. 
\ref{subsec.single}.  All other values were either calculated or 
extracted from direct measurements as described in this section.}
\label{tab.char}
\end{center}
\end{table}

For canonical hydrogen maser operation, there are two important 
relaxation rates \cite{Hmas2,ed.ieee}.  For a room temperature H 
maser, the decay of the population inversion is described 
by the longitudinal relaxation rate
\begin{equation}
    \label{eqn.gamma1}
    \gamma_{1} = \gamma_{b} + \gamma_{r} +  2\gamma_{se} + \gamma_{1}^{\prime},
\end{equation}
and the decay of the atomic coherence is described by 
the transverse relaxation 
rate 
\begin{equation}
    \label{eqn.gamma2}
    \gamma_{2} = \gamma_{b} + \gamma_{r} +  \gamma_{se} + \gamma_{2}^{\prime}.
\end{equation}
Here, $\gamma_{b}$ is the atomic flow rate into the bulb, 
$\gamma_{r}$ is the rate of recombination into molecular hydrogen at 
the bulb wall, $\gamma_{se}$ is the hydrogen-hydrogen spin-exchange 
decay rate, and $\gamma_{i}^{\prime}$ includes all other sources of 
decay, such as decoherence during wall collisions and effects of 
magnetic field gradients.  

In the steady state, the atom flow rate into the bulb is equal to the 
geometric escape rate from the bulb, given by $\gamma_{b} = \bar{v} 
A / 4 K V_{b}$, where $\bar{v}$ =  2.5 $\times 10^{5}$ cm/s is the mean thermal velocity 
of atoms in the bulb, $A$ = 0.254 cm$^{3}$ is the area of the 
bulb entrance aperture, and $K \approx$ 6 is the Klausing factor \cite{mol.beams}.  
Thus, $\gamma_{b}$ = 0.86 rad/s for both P-8 and P-28.
The spin exchange decay rate is given approximately by \cite{Hmas2,ed.ieee}
\begin{equation}
    \label{eqn.gammase}
    \gamma_{se} = \frac{1}{2} n \bar{v}_{r} \sigma 
\end{equation}
where $\bar{v}_{r}$ = 3.6 $\times 10^{5}$ cm/s is the mean relative velocity of atoms in the 
bulb and $\sigma$ = 21 $\times$ 10$^{-16}$ cm$^{2}$ is the 
hydrogen-hydrogen spin-exchange cross section.  The hydrogen density 
is given by \cite{Hmas2,ed.ieee}
\begin{equation}
    \label{eqn.density}
    n = \frac{I_{tot}}{(\gamma_{b}+\gamma_{r})V_{b}}
\end{equation}
where $I_{tot}$ is the total flux of hydrogen atoms into the storage 
bulb.  

The atomic line-Q is related to the transverse relaxation rate and the maser oscillation 
frequency $\omega$ by \cite{Hmas1,ed.ieee}
\begin{equation} 
    \label{eqn.lineQ}
    Q_{l} = \frac{\omega}{2\gamma_{2}}.  
\end{equation}    
It is measured using the cavity pulling of the maser 
frequency:  neglecting spin-exchange shifts, the maser frequency is given 
by \cite{Hmas1}
  \begin{equation}
  \label{eqn.cavpull}
      \omega = \omega_{24} + \frac{Q_{C}}{Q_{l}} \left( \omega_{C} - 
      \omega_{24} \right).
  \end{equation}
By measuring the maser frequency as a function of cavity frequency setting, 
the line-Q can be 
determined.  For both P-8 and P-28, we find $Q_{l}$ = 1.6 $\times$ 
10$^{9}$, and therefore $\gamma_{2}$ = 2.8 rad/s.

A convenient single measure of spin-exchange-independent 
relaxation in a hydrogen maser is given by ``gamma-t'' \cite{Hmas2,ed.ieee}
\begin{equation}
    \label{eqn.gammat}
    \gamma_{t} = \left[ 
    (\gamma_{b} + \gamma_{r} + \gamma_{1}^{\prime}) 
    (\gamma_{b} + \gamma_{r} + \gamma_{2}^{\prime}) \right] ^{\frac{1}{2}}.
\end{equation} 
Using this, a more useful form for the longitudinal relaxation rate, $\gamma_{1}$, can be 
found.  By combining Eqn. \ref{eqn.gammat} with Eqns. \ref{eqn.gamma1} and \ref{eqn.gamma2}, 
we find 
\begin{equation}
    \label{eqn.gamma1exp}
    \gamma_{1} = \frac{\gamma_{t}^{2}}{\gamma_{2} - \gamma_{se}} + 2 
    \gamma_{se}.
\end{equation}

Using Eqns. \ref{eqn.gamma2}-\ref{eqn.lineQ}, we can relate the line-Q to 
$I$, the input flux of atoms in state $|2\rangle$ as \cite{ed.ieee}
\begin{equation}
    \label{eqn.lineQ2}
    \frac{1}{Q_{l}} = \frac{2}{\omega}
    \left[\gamma_{b}+\gamma_{r}+\gamma_{2}^{\prime}+q\frac{I}{I_{th}}\gamma_{t} \right]
\end{equation}
using the threshold flux required for maser oscillation (neglecting 
spin-exchange)
\begin{equation}
    \label{eqn.ith}
       I_{th} = \frac{\hbar V_{C} \gamma_{t}^{2}}{4 \pi \mu_{B}^{2} 
       Q_{C} \eta},
\end{equation}
and the maser quality parameter
\begin{equation}
    \label{eqn.q}
      q = \left[ \frac{\sigma \bar{v}_{r} \hbar}{8 \pi \mu_{b}^{2}} \right]
           \frac{\gamma_{t}}{\gamma_{b}+\gamma_{r}}
	   \left[ \frac{V_{C}}{\eta V_{b}} \right]
	   \left( \frac{1}{Q_{C}} \right)
	   \frac{I_{tot}}{I}.
\end{equation}

The ratio $I$/$I_{tot}$ is a measure of the 
effectiveness of the state selection of atoms entering the bulb.  
While $I$ is not directly measurable, it can be related to the power 
$P$ radiated by the atoms by \cite{Hmas2,ed.ieee}
\begin{equation}
    \label{eqn.ppc}
    \frac{P}{P_{c}} = -2q^{2} \left( \frac{I}{I_{th}} \right)^{2} + 
    (1-3q) \left( \frac{I}{I_{th}} \right) -1
\end{equation}
where $P_{c} = \hbar \omega I_{th} / 2$.
The maser power is also related to the maser Rabi frequency by 
\cite{Hmas2}
\begin{equation}
    \label{eqn.maser.power}
    P = \frac{I \hbar \omega}{2} \frac{|X_{24}|^{2}}{\gamma_{1} \gamma_{2}} 
        \left( 1+\frac{|X_{24}|^{2}}{\gamma_{1} 
        \gamma_{2}} \right)^{-1}.
\end{equation}
The power coupled out of the maser is given by \cite{ed.ieee}
$P_{o}/P =  \beta / (1+\beta)$.

Generally, the parameter $q$ is less than 0.1, while $I/I_{th}$ is 
approximately 2 or 3.  Hence, the first term of Eqn. \ref{eqn.ppc} can 
be neglected relative to the others.  If we make the reasonable 
approximation that $\gamma_{1}^{\prime}$ = $\gamma_{2}^{\prime}$, then we 
can rewrite Eqn. \ref{eqn.lineQ2} using Eqns. \ref{eqn.gammat}, 
\ref{eqn.ith}-\ref{eqn.ppc} as \cite{ed.ieee}
\begin{equation}
    \label{eqn.lineQ3}
    \frac{1}{Q_{l}} = mP+b
\end{equation}
where
\begin{equation}
    \label{eqn.b}
    b = \frac{2}{\omega} \gamma_{t} \left[ 1 + \frac{q}{1-3q} \right]
\end{equation}
and
\begin{equation}
    \label{eqn.m}
    m = \frac{16 \pi \mu_{b}^{2} Q_{C} \eta}{\omega^{2} \hbar^{2} V_{C}}
         \left[ \frac{q}{1-3q} \right] \frac{1}{\gamma_{t}}.
\end{equation}
Therefore, by measuring the line-Q as a function of maser power and 
extracting the slope $m$ and the y-intercept $b$, we can determine $q$ and 
$\gamma_{t}$.  For maser P-8, $q$ = 0.100 and $\gamma_{t}$ = 1.83 rad/s.

With these values of $q$ and $\gamma_{t}$, we found $I_{th}$ = 0.54 $\times$ 10$^{12}$ 
atoms/s (using Eqn. \ref{eqn.ith}), and $P_{c}$ = 250 fW.  With a 
measured output power of $P_{o}$ = 112 fW, the atoms were radiating 
$P$ = 599 fW, and the flux of state $|2\rangle$ atoms was
$I$ = 3.13 $\times$ 10$^{12}$ atoms/s (Eqn. \ref{eqn.ppc}).  Under the assumption 
that $\gamma_{t} \approx \gamma_{b} + \gamma_{r}$ we found that the total 
flux was $I_{tot}$ = 15.0 $\times$ 10$^{12}$ atoms/s (Eqn. \ref{eqn.q}) and the density 
was $n$ = 2.8 $\times$ 10$^{15}$ atoms/m$^{3}$ (Eqn. \ref{eqn.density}).  
The spin-exchange decay rate was then found to be $\gamma_{se}$ = 1.06 
rad/s (Eqn. \ref{eqn.gammase}).
Finally, the population decay rate was $\gamma_{1}$ = 4.04 rad/s (Eqn. \ref{eqn.gamma1exp}) 
and the maser Rabi frequency was $|X_{24}|$ = 2.77 rad/s (Eqn. 
\ref{eqn.maser.power}).

\subsection{Zeeman frequency determination}
\label{subsec.zeeman}

\begin{figure}
\begin{center}
\includegraphics{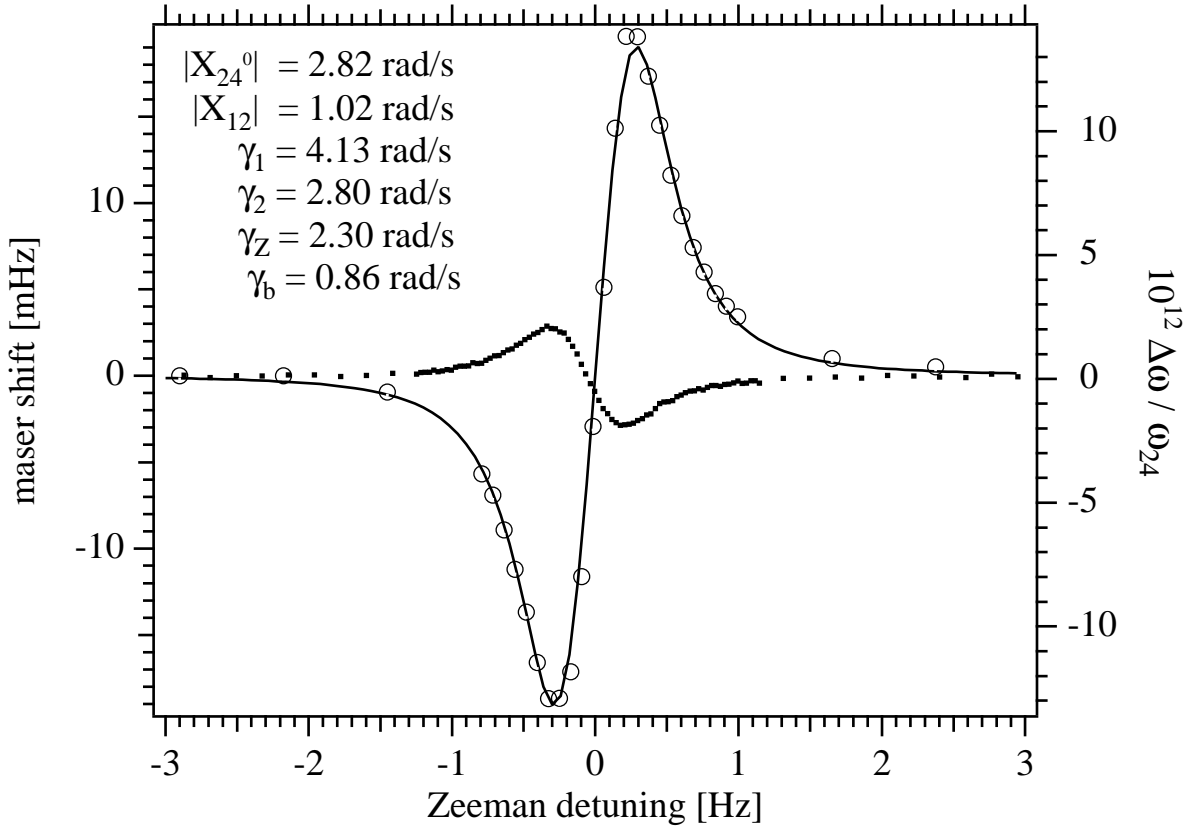}
\caption{Double resonance maser frequency shifts.  The large open 
  circles (maser P-8) are compared with
  Eqn.~\ref{eqn.analytic.andresen} (full curve) using the parameter
  values shown.  The values of $|X_{12}|$ and $\gamma_{Z}$ were chosen
  to fit the data, while the remaining parameters were independently
  measured as described in subsection \ref{subsec.numbers}.  The
  experimental error of each measurement (about 40 $\mu$Hz) is smaller
  than the circle marking it.  The solid square data points are data from
  the Lorentz symmetry test (maser P-28).  The large variation of
  maser frequency with Zeeman detuning near resonance, along with the
  excellent maser frequency stability, allows the Zeeman frequency
  ($\approx$ 800 Hz) to be determined to 3 mHz in a single resonance
  (requiring 18 minutes of data acquisition).  The inversion of the
  shift between the two is due to the fact that for the P-8 data 
  (open circles), the
  maser operated with an input flux of $|2\rangle$ and $|3\rangle$
  atoms, while for the P-28 data (solid points), the typical input of $|1\rangle$ and
  $|2\rangle$ atoms was used.  Changing between these two input flux 
  modes is done by inverting the direction of the static solenoid 
  field, while maintaining a fixed quantization axis for the state 
  selecting hexapole magnet (see Sec. \ref{subsec.runs2,3}).}
\label{fig.andresen}
\end{center}
\end{figure}

The $F=1$, $\Delta m_{F} = \pm 1$ Zeeman frequency is measured using a
double resonance technique \cite{andresen,savard,doubres}.  As the
frequency of an audio frequency magnetic field $\omega_{Z}$, applied
perpendicular to the quantization axis, is swept through the
$\emph{Zeeman}$ frequency, a shift in the $\emph{maser}$ frequency
is observed (Fig.~\ref{fig.andresen}).  When the applied field is near
the Zeeman frequency, two-photon transitions (one audio photon plus
one microwave photon) link states $|1\rangle$ and $|3\rangle$ to state
$|4\rangle$, in addition to the single microwave photon transition
between states $|2\rangle$ and $|4\rangle$.  This two photon coupling
shifts the maser frequency antisymmetrically with respect to the
detuning of the applied field about the Zeeman resonance \cite{doubres}.

To second order 
in the Rabi frequency of the applied Zeeman field, $|X_{12}|$
the small static-field limit of the maser frequency 
shift from the unperturbed frequency is given by \cite{andresen}
  \begin{eqnarray}
    \label{eqn.analytic.andresen}
    \Delta \omega & = & -|X_{12}|^{2} (\rho_{11}^{0} - \rho_{33}^{0}) 
    \frac{\delta (\gamma_{1} 
    \gamma_{2}+|X_{24}^{0}|^{2})(\gamma_{Z}/\gamma_{b})}
    {(\gamma_{Z}^{2} - \delta^{2} + \frac{1}{4} |X_{24}^{0}|^{2})^{2} + 
    (2 \delta \gamma_{Z})^{2}} \\
    & & + |X_{12}|^{2} \left( \frac{\omega_{C} - \omega_{24}}{\omega_{24}} \right)
    \frac{Q_{C}\gamma_{Z}(1+K)}{\gamma_{Z}^{2}(1+K)^{2}+\delta^{2}(1-K)^{2}}  \nonumber
  \end{eqnarray} 
where $\gamma_{Z}$ is the Zeeman decoherence rate, $\delta = \omega_{Z} - \omega_{23}$ 
is the detuning of the applied field from 
the atomic Zeeman frequency, $K = \frac{1}{4} |X_{24}^{0}|^{2} / (\gamma_{Z}^{2}+ 
\delta^{2})$, and $\rho_{11}^{0} - \rho_{33}^{0} = \gamma_{b} / (2 \gamma_{1})$ is the steady 
state population difference between states $|1\rangle$ and $|3\rangle$ in the absence 
of the applied Zeeman field.  
The first term in Eqn.~\ref{eqn.analytic.andresen} results from the 
coherent two-photon mixing of the $F=1$ levels as described above \cite{doubres}, 
while the second term is a modified cavity pulling term that results 
from the reduced line-Q in the presence of the applied Zeeman field.  
We compared Eqn.~\ref{eqn.analytic.andresen} to experimental data 
from P-8, inserting the independently measured values of $|X_{24}^{0}|$, 
$\gamma_{b}$, $\gamma_{1}$, and $\gamma_{2}$.  By matching the fit to the data 
we extracted the Zeeman field parameters $|X_{12}|$ and 
$\gamma_{Z}$ shown in Fig.~\ref{fig.andresen}.

In addition to the shift given by Eqn. \ref{eqn.analytic.andresen}, there is a small 
symmetric frequency shift due to the slight non-degeneracy of the two 
$F=1$, $\Delta m_{F} = \pm 1$ Zeeman frequencies.  This  
term offsets the zero crossing of the maser shift 
resonance away from the 
average Zeeman frequency $\frac{1}{2} \left( \nu_{12} + 
\nu_{23} \right)$, however the contribution is negligible at small static fields.  
Also, a reanalysis of 
the double resonance maser shift \cite{savard}, which included the 
effects of spin-exchange collisions \cite{bender}, showed that there is 
an additional hydrogen density-dependent offset of the zero crossing 
of the maser shift resonance from the average Zeeman frequency.  
Using the full spin-exchange corrected formula for the maser 
frequency shift \cite{savard}, we calculated this
offset and found that for typical hydrogen maser 
densities ($n \approx$ 3 $\times 10^{15}$ m$^{-3}$), the offset varied with
average maser power as approximately -50 $\mu$Hz/fW (assuming a 
linear relation between maser power and atomic density of 
$\frac{\Delta P}{\Delta n} \approx \frac{100 \ fW}{3 \times 10^{15} \ 
m^{-3}}$).  As described 
below, our masers typically have sidereal power fluctuations less than 1 fW, 
making this effect negligible.

\begin{figure}
\begin{center}
\includegraphics{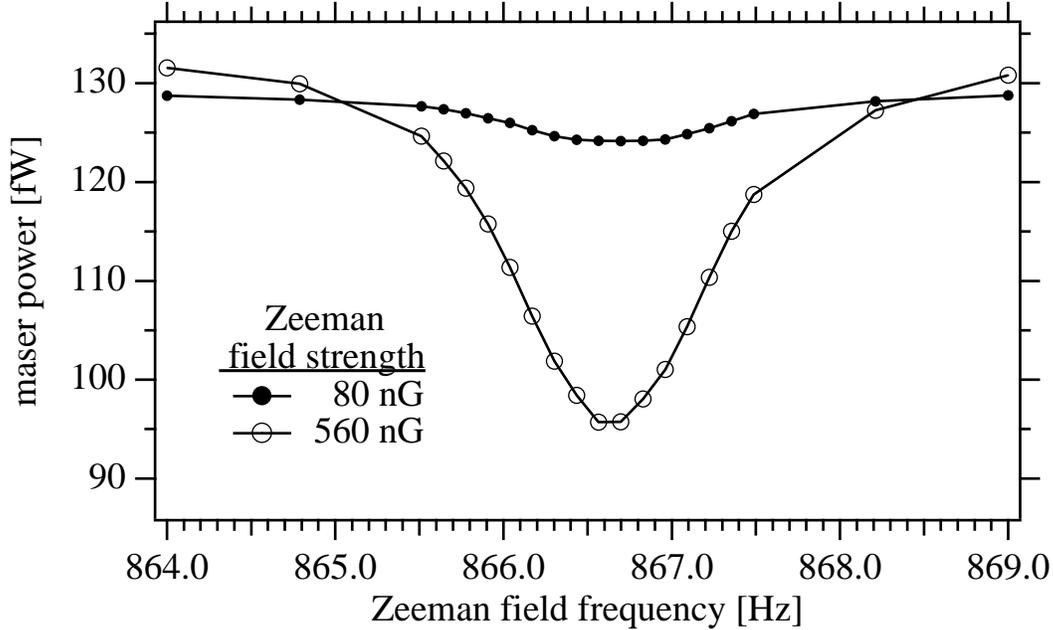}
\caption{Double resonance maser power diminishment.  The open circles, taken 
with an applied Zeeman field strength of about 560 nG, represent 
typical data used to determine the value of the static magnetic field 
in the maser bulb.  The filled circles are maser power 
curves with an applied field strength of about 80 nG.  Our Lorentz 
symmetry test data were taken 
with a field strength of about 35 nG, where the power diminishment is 
less than 2\%.}
\label{fig.powerdips}
\end{center}
\end{figure}

The applied Zeeman field also acts to diminish the maser power, as
shown in Fig.~\ref{fig.powerdips}, and to decrease the maser's line-Q.
By driving the $F = 1$, $\Delta m_{F} = \pm 1$ Zeeman transitions, the
applied field depletes the population of the upper masing state
$|2\rangle$, thereby diminishing the number of atoms undergoing the
maser transition and reducing the maser power.  Also, by decreasing
the lifetime of atoms in state $|2\rangle$, the line-Q is
reduced.  A very weak Zeeman field of about 35 nG (as was used in
our Lorentz symmetry test) decreases the maser power by less
than 2\% on resonance and reduces the line-Q by 2\% (as calculated
using Eqn.\ 6 of \cite{andresen}).  The standard method of determining
the average static magnetic field strength is to scan the Zeeman resonance
with a large applied field and record the power diminishment (such as
that shown in Fig.~\ref{fig.powerdips}, open circles).  From the
applied field frequency at the center of the power resonance, which
typically has a width of about 1 Hz, the magnetic field can be found
with a resolution of about 1 $\mu$G.

\section{Experimental procedure}
\label{sec.experiment}

\subsection{Zeeman frequency measurement}
\label{subsec.single}

To measure the $F=1$, $\Delta m_{F}=\pm1$ Zeeman frequency, we applied an oscillating field
of about 35 nG near the Zeeman frequency. 
This field shifted the maser frequency by a few mHz (at the extrema), a fractional 
shift of about 2 parts per trillion.  
Because of the excellent fractional maser stability (2 parts in 
$10^{14}$ over our averaging times of 10 s), 
the shift was easily resolved (see the solid data in Fig. 
\ref{fig.andresen}).  As the frequency of the applied field 
was stepped through the Zeeman resonance, 
the maser frequency (of perturbed maser P-28) was compared to a second, 
unperturbed hydrogen maser frequency (P-13).  The two maser signals 
at $\approx$ 1420 MHz were phase locked to independent voltage controlled crystal 
oscillator receivers.  The exact value of the receivers' outputs 
were set by tunable synthesizers, which were set such that there 
was a 1.2 Hz offset between them.  The two receiver outputs were combined 
in a heterodyne mixer and the resulting 0.8 s period beat note
was averaged for 10 s (about 12 periods) with a Hewlett-Packard Model HP 5334B
frequency counter.  The full double resonance spectrum consisted of 100 
such points.  For each spectrum, 80\% of the points were 
taken over the middle 40\% of the scan range, where the frequency 
shift varies the most.     

Once an entire spectrum of beat period vs applied Zeeman frequency was obtained, 
it was fit to the function
\begin{eqnarray}
    \label{eqn.fit.function}
    T_{b} & = A_{0} & + \frac{A_{3} \delta (1-\kappa)}{A_{1} (1+\kappa)^{2} + \delta^{2} (1-\kappa)^{2}}
                      - \frac{A_{3} (\delta + \tau) (1-\kappa)}{A_{1} (1+\kappa)^{2} + 
                                                                     (\delta+\tau)^{2} (1-\kappa)^{2}} \\
    & & + \frac{A_{5} \delta}{(A_{1} - \delta^{2} + A_{4})^{2} + 4\delta^{2} A_{1}} 
        + \frac{A_{6} (1+\kappa)}{A_{1} (1+\kappa)^{2} + \delta^{2} (1-\kappa)^{2}} \nonumber 
\end{eqnarray}
to determine the Zeeman frequency.  Here $\delta = \nu - \nu_{Z}$ is
the Zeeman detuning of the applied field $\nu$ away from the Zeeman
frequency $\nu_{Z}$, $\kappa = A_{4} / (A_{1} + \delta^{2})$ is the
analog of the parameter $K$ from Eqn.~\ref{eqn.analytic.andresen}, and
$\tau = (1.403 \times 10^{-9}) \times \nu_{Z}^{2}$ is the small
difference between the two Zeeman frequencies $\nu_{12}$ and
$\nu_{23}$.  The first term $A_{0}$ is the constant offset
representing the unperturbed beat period between the two masers.  The
second and third terms comprise the first-order symmetric maser shift
(not included in Eqn.~\ref{eqn.analytic.andresen} but described in the
text above); these two terms nearly cancel at low static field where
$\tau$ vanishes.  The final two terms account for the two shifts given
in Eqn.~\ref{eqn.analytic.andresen}.

For our spectra in maser P-28 with small 
applied field amplitude (solid square data points of Fig. 
\ref{fig.andresen}), typical fit parameters were:  
$A_{0}$ = 0.84550 $\pm$ 0.00001, $A_{1}$ = 0.141 $\pm$ 0.005, 
$\nu_{Z}$ = 857.063 $\pm$ 0.003, $A_{3}$ = 0.006 $\pm$ 0.010, $A_{4}$ = 
0.029 $\pm$ 0.003, $A_{5}$ = (3.2 $\pm$ 0.1) $\times 10^{-4}$, and $A_{6}$ = 
(-1 $\pm$ 5) $\times 10^{-6}$.  The uncertainty in the 
Zeeman frequency was 3 mHz.  Also, $A_{3}$ and $A_{6}$, the amplitude 
coefficients of the residual first order effect and the cavity 
pulling term, were consistent with zero. 

With our known value of $\gamma_{b}$ = 0.86 rad/s, and our measured value of 
$\gamma_{2}$ = 2.77 rad/s (from line-Q), the above set of fit parameters 
were consistent with the reasonable values $X_{24}^{0}$ = 2.14 rad/s, 
$\gamma_{Z}$ = 2.36 rad/s, $\gamma_{1}$ = 2.88 rad/s, and 
$X_{12}$ = 0.40 rad/s (since $A_{3}$ had such a large error bar, the value 
of $X_{12}$ was chosen such the ratio of the maser shift amplitude in P-28 to P-8, 
shown in Fig.~\ref{fig.andresen}, is equal to the ratio of the
squares of $X_{12}$ for P-28 to P-8). 

\begin{figure}
\begin{center}
\includegraphics{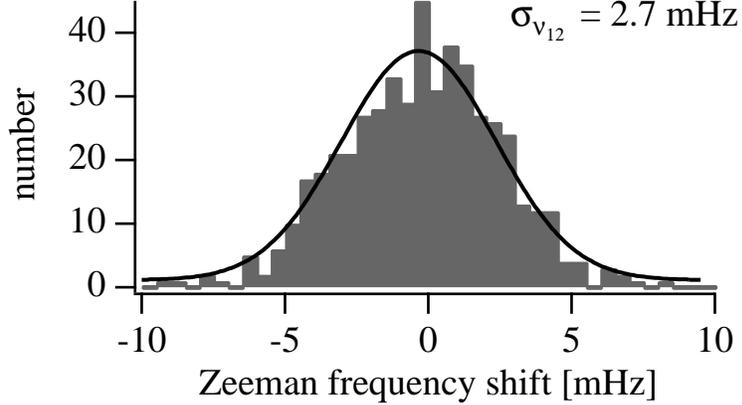}
\caption{Results from a Monte Carlo analysis.  The horizontal axis 
represents the shift of the Zeeman frequency as determined by our fits
of over 100 synthetic data sets, the vertical axis is the number within 
each shift bin.  The width of the Gaussian fit to the data is 2.7 
mHz, representing the resolution of a single Zeeman frequency 
measurement.}
\label{fig.montecarlo}
\end{center}
\end{figure}

To determine the number of points and length of averaging that
optimized the Zeeman frequency resolution, we recorded several spectra
with 50, 100, and 150 points at 5 s and 10 s averaging.  We also
varied the ``density distribution'' of points, including spectra where
the middle 40\% of the scan contained 80\% of the points and those
where the middle 30\% contained 80\% of the points (thus increasing
the number of points in the region where the antisymmetric shift
varies the most).  With each of these spectra, we ran the following
Monte Carlo analysis \cite{num.rec.montecarlo}: after fitting each
scan to Eqn.~\ref{eqn.fit.function}, we constructed 100 synthetic data
sets by adding Gaussian noise to the fit, with noise amplitude
determined by the unperturbed maser frequency resolution of about 40
$\mu$Hz.  Each of these synthetic data sets was fit and a histogram of
the fitted Zeeman frequencies was constructed.  The resolution of each
spectrum was taken as the width of the Gaussian curve that fit the
histogram (see Fig.~\ref{fig.montecarlo}).  As the total
length of the scans increased, the resolution improved and converged
to a limit of around 2.5 mHz.  While the resolution improved slowly
with increased acquisition time, it would have eventually begun to
degrade due to long term drifting of the Zeeman frequency.  (As will
be described below, we found that the Zeeman frequency exhibited slow
drifts of about 10-100 mHz/day).  We therefore chose a scan of 100
points at 10 s averaging, for a total length of about 18 minutes for
our Lorentz symmetry test spectra.  The results from the Monte Carlo analysis
for one of these spectra indicated a Zeeman frequency resolution is
2.7 mHz (see Fig.~\ref{fig.montecarlo}).

\subsection{Data analysis}
\label{subsec.analysis}

Our net result combines data from three runs.  During each data run, 
the 18 minute Zeeman frequency scans were automated and run 
consecutively.  After every 10 scans, 20 minutes of ``unperturbed'' 
maser frequency stability data was taken to track the 
maser's stability.  Each run contained about 10 continuous days worth of 
data, and each set contained more than 500 Zeeman frequency 
measurements, taken at $\approx$ 18 minute intervals.
  
For each run, the long term Zeeman frequency data was fit to a function of the form
\begin{equation}
    \label{eqn.fit}
    fit\ =\ (\mathit{piecewise\ continuous\ linear\ function}) + \delta\nu_{Z,\alpha} 
\cos(\omega_{sid}t) + \delta\nu_{Z,\beta} \sin(\omega_{sid}t)
\end{equation}    
where $\delta\nu_{Z,\alpha}$ and $\delta\nu_{Z,\beta}$ represent the cosine and 
sine components of the sidereal sinusoid.  The time origin of
the sinusoids for all three runs was taken as midnight (00:00) of November 19, 1999.  The 
subscripts $\alpha$ and $\beta$ refer to two non-rotating orthogonal axes 
perpendicular to the rotation axis of the earth.  The 
total sidereal amplitude was determined by adding $\delta\nu_{Z,\alpha}$ 
and $\delta\nu_{Z,\beta}$ in quadrature.  During each run, the Zeeman frequency 
drifted hundreds of mHz over tens of days.  
The piecewise continuous linear function, consisting of 
segments one sidereal day in length, was included to account for these long 
term Zeeman frequency drifts.  This function was continuous at 
each break, while the derivative was discontinuous.  

The result of this analysis, where the fitting function
(Eqn.~\ref{eqn.fit}) was applied to the full data set, was found to be
in good agreement with a second analysis, where each individual day of
data was fit to a line plus the sidereal sinusoid and the
cosine and sine amplitudes of each day were averaged separately and
then combined in quadrature to find the total sidereal amplitude.

\subsection{Run 1}
\label{subsec.run1}

\begin{figure}
\begin{center}
\includegraphics{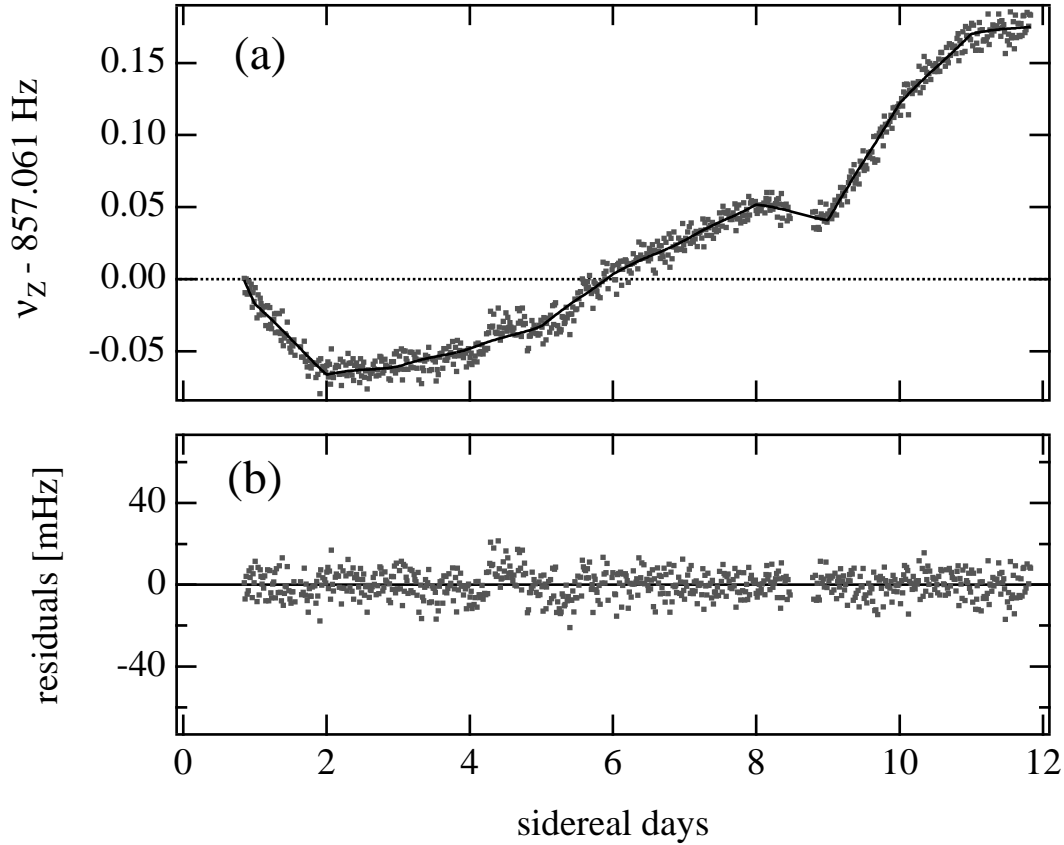}
\caption{(a) Run 1 data (November 1999), with solenoid current fluctuations 
subtracted.  From the measured Zeeman frequencies, we subtracted 
857.061 Hz.  (b) Residuals after fitting the data to Eqn.~\ref{eqn.fit}.}
\label{fig.run1}
\end{center}
\end{figure}

The cumulative data from the first run (November 1999) are shown in 
Fig.~\ref{fig.run1}(a) 
and the residuals from the complete fit (Eqn.~\ref{eqn.fit}) are shown 
in Fig.~\ref{fig.run1}(b).  The data  set consisted of 11 full days of data 
and had an overall drift of about 250 mHz.  

\begin{figure}
\begin{center}
\includegraphics{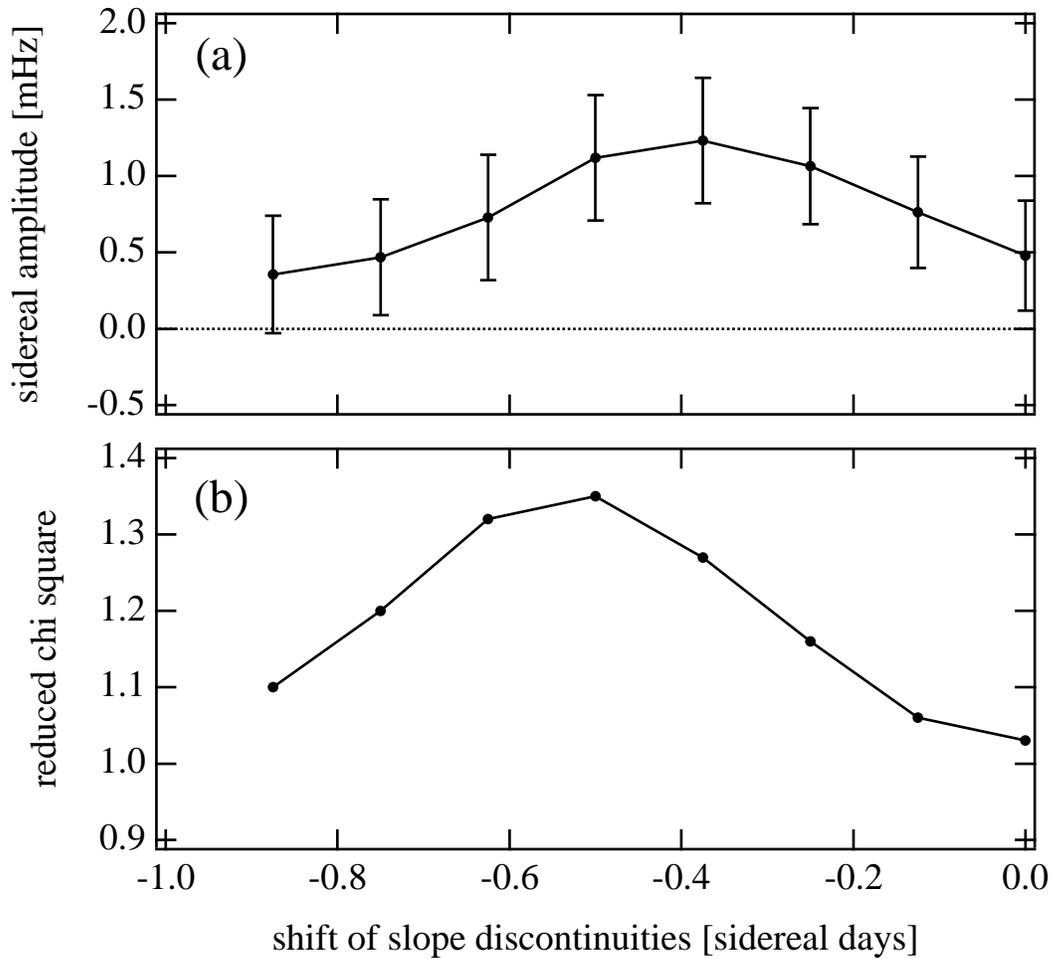}
\caption{(a) Total sidereal amplitudes for the first run.  The different points 
are from different choices of slope discontinuity locations.  (b) 
Corresponding reduced chi square parameters.  The minimum value occurs 
with a slope break origin of midnight (00:00) of November 19, 1999.}
\label{fig.run1shift}
\end{center}
\end{figure}

To avoid a biased choice of fitting, we allowed the location of the 
slope discontinuities in the piecewise continuous linear function to shift throughout a sidereal day.  
We made eight separate fits, each with the location of the slope 
discontinuities shifted by three sidereal hours.  The total sidereal amplitude and 
reduced chi square for each is shown in Fig.~\ref{fig.run1shift}.  
We chose our result from the fit with minimum reduced chi square.

As noted above, the error bar on a single Zeeman frequency 
determination was about 3 mHz.  However, when analyzing a smooth 
region of long term Zeeman data (about 1 day) we calculate a
standard deviation of about 5 mHz.  We 
believe this error bar is due mainly to residual thermal 
fluctuations (see Fig. \ref{fig.temp.data}).

For our choice of slope discontinuity with minimum reduced chi square \cite{maxchi}, the 
cosine amplitude was 0.43 mHz $\pm$ 0.36 mHz, and the sine amplitude was 
-0.21 mHz $\pm$ 0.36 mHz.  The total sidereal amplitude was therefore 0.48 
mHz $\pm$ 0.36 mHz.  

\subsection{Field-inverted runs 2 and 3}
\label{subsec.runs2,3}

In runs 2 and 3, the static solenoid field orientation was opposite that 
of the initial run to further study the double resonance technique and 
any potential systematics associated with the solenoid field.
With the static field inverted, and therefore directed opposite the 
quantization axis in the state selecting hexapole magnet, the input 
flux consists of atoms in states $|2\rangle$ and $|3\rangle$ (rather 
than the states $|1\rangle$ and $|2\rangle$).  Thus, 
reversing the field inverts the steady state population difference $(\rho_{11}^{0} - 
\rho_{33}^{0})$ of Eqn. \ref{eqn.analytic.andresen} and acts to 
invert the antisymmetric double resonance maser frequency shift 
\cite{doubres}.

\begin{figure}
\begin{center}
\includegraphics{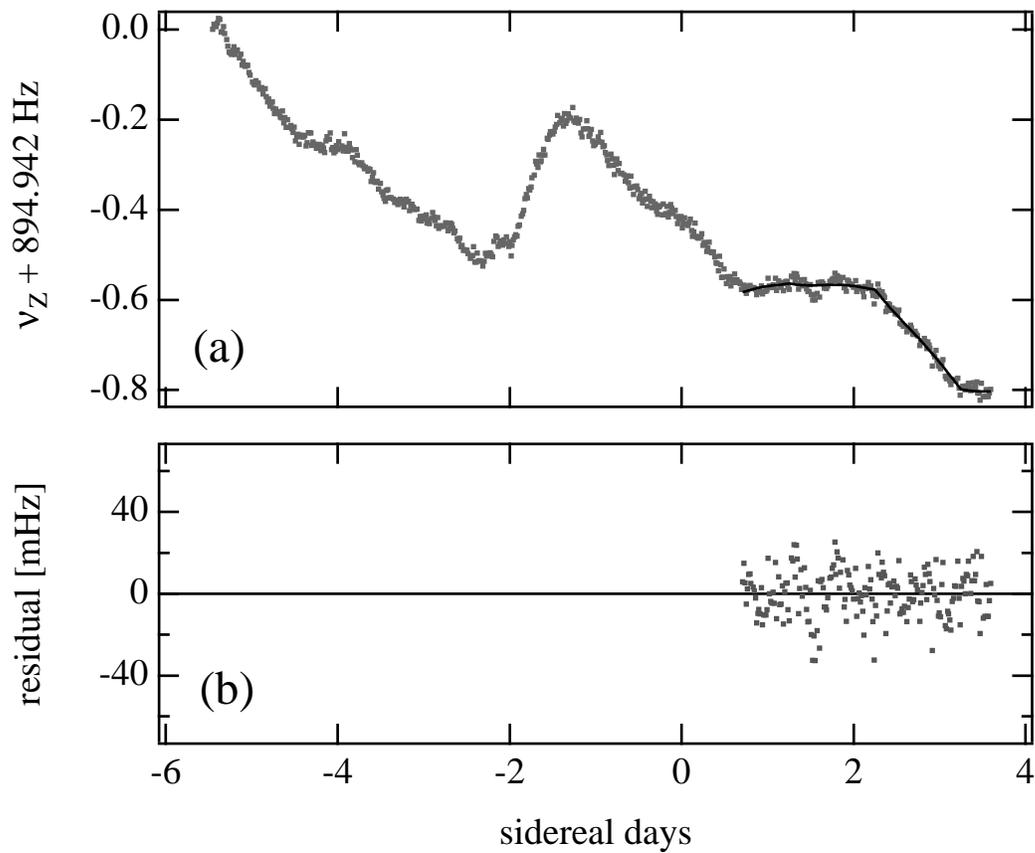}
\caption{(a) Run 2 data (December 1999), with solenoid current fluctuations 
  subtracted.  To the measured Zeeman frequencies, we added 894.942
  Hz.  (Note the sign reversal from run 1 to account for the inverted
  field).  (b) residuals after fitting the data to
  Eqn.~\ref{eqn.fit}.}
\label{fig.run2}
\end{center}
\end{figure}

\begin{figure}
\begin{center}
\includegraphics{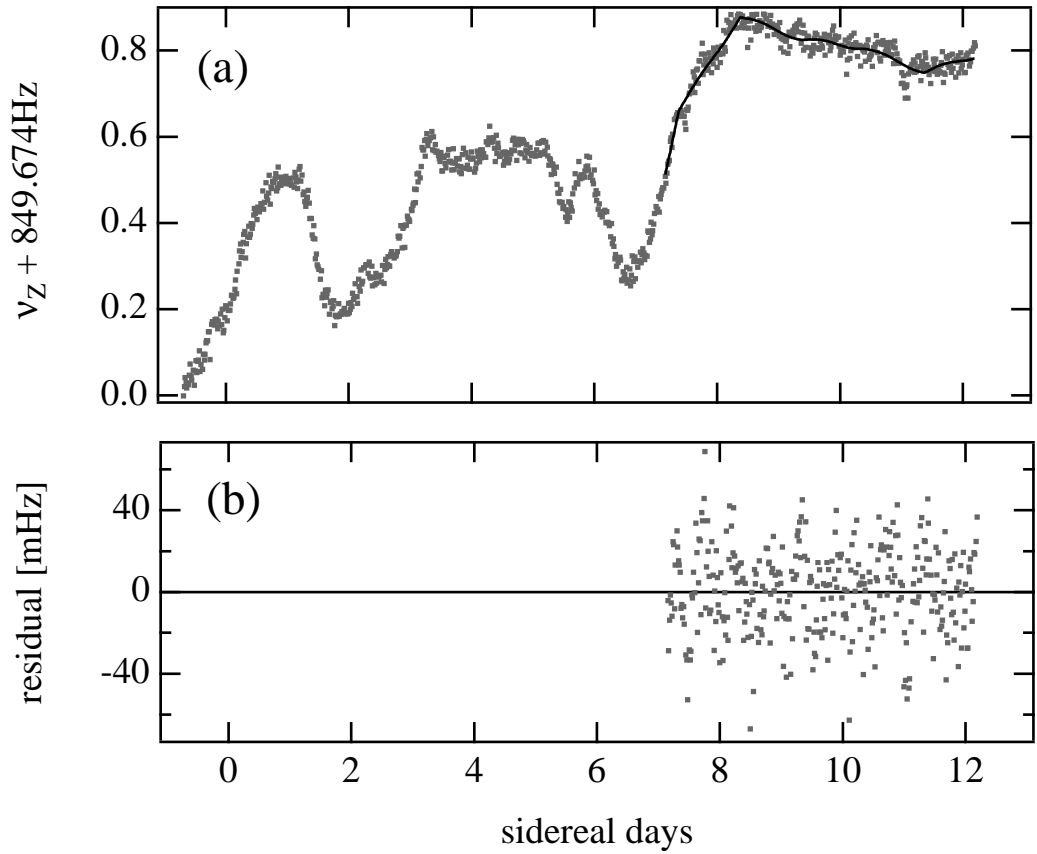}
\caption{(a) Run 3 data (March 2000), with solenoid current fluctuations 
  subtracted.  To the measured Zeeman frequencies, we added 849.674
  Hz.  (Note the sign reversal from run 1 to account for the inverted
  field).  (b) residuals after fitting the data to
  Eqn.~\ref{eqn.fit}.}
\label{fig.run3}
\end{center}
\end{figure}

Operating the maser in the field 
reversed mode degrades the maser performance and subsequently the 
Zeeman frequency data.  With opposed quantization fields inside the maser
bulb and at the exit of the state selecting hexapole magnet, a narrow
region of field inversion is created.  Where the field passes through zero,
Majorana transitions between the different $m_{F}$ sublevels of the
$F=1$ manifold can occur.  This can alter the number of atoms in the
upper maser state ($F=1$, $m_{F}=0$, state $|2\rangle$), which diminishes the overall
maser amplitude and stability.  In the field-inverted configuration,
the maser amplitude was reduced by 30\%, and both the maser frequency
and Zeeman frequency were less stable.  In addition, the
field-inverted runs were each conducted soon after a number of rather
invasive repairs were made to the maser \cite{P8/P28}.  Thus, the quality of the
latter two data sets was somewhat degraded from the first run (see 
Figs.~\ref{fig.run2}(a) and \ref{fig.run3}(a)).  The
overall drift was larger (nearly 800 mHz over about 10 days), and the
scatter in the data was increased, as can be seen from the residual
plots from these runs (Figs.~\ref{fig.run2}(b) and \ref{fig.run3}(b))
which have been plotted on the same scale as the residuals from the
first run (Fig.~\ref{fig.run1}(b)).

The latter two runs were also less suitable for the piecewise 
continuous linear drift model 
used in the first run.  In that case, the large slope 
changes were coincidentally separated by an integer number of sidereal 
days; in the last two runs, the larger and more frequent changes in 
slope were not.  
Therefore, only certain selected sections could be fit to the same 
model (Eqn.~\ref{eqn.fit}), significantly truncating the data sets.  
Due to all of these factors, the sidereal amplitudes and the associated 
error bars were up to an order of magnitude larger for the 
field-inverted runs than the first run.  All values are shown together in Table 
\ref{tab.results}.
    
\begin{table}
\begin{center}
\begin{tabular}{c c c c c}
    Run  & $\delta\nu_{Z,\alpha}$ [mHz] & $\sigma_{\alpha}$ [mHz]& 
$\delta\nu_{Z,\beta}$ [mHz]& $\sigma_{\beta}$ [mHz] \\ 
    \hline
    1 & 0.43 & 0.36 & -0.21 & 0.36 \\ 
    2 & -2.02 & 1.27 & -2.75 & 1.41 \\
    3 & 4.30 & 1.86 & 1.70 & 1.94 \\
\end{tabular}   
\caption{Sidereal amplitudes from all runs.}
\label{tab.results}
\end{center}
\end{table}

\subsection{Combined result}
\label{subsec.ave}

The final sidereal bound, combining all three runs, was calculated 
using the data in Table \ref{tab.results}.  First, the weighted averages 
of the cosine and sine amplitudes, $\bar{\delta \nu}_{Z,\alpha}$ and $\bar{\delta 
\nu}_{Z,\beta}$, were found using the standard formula for weighted 
mean \cite{bev.weight}
\begin{equation}
    \mu^{\prime} = \left( \Sigma\frac{x_{i}}{\sigma_{i}^{2}} 
    \right) /  \left(\Sigma\frac{1}{\sigma_{i}^{2}} \right),
\end{equation} 
and their uncertainties were given by
\begin{equation}
    \sigma_{\mu^{\prime}}^{2} = \sqrt{ 1 / \left( 
    \Sigma\frac{1}{\sigma_{i}^{2}} \right)}.
\end{equation}
The sign reversal due to the field inversion was accounted 
for in the raw data, before the data were fit.  Thus, the runs are 
combined using conventional (i.e., additive) averaging.
The final sidereal amplitude $A$ was calculated by adding the mean cosine 
and sine amplitudes in quadrature, $A = \sqrt{\bar{\delta 
\nu}_{Z,\alpha}^{2} + \bar{\delta \nu}_{Z,\beta}^{2}}$.
We measure a sidereal variation of the $F=1$, $\Delta m_{F}=\pm 
1$ Zeeman frequency of hydrogen of $A = 0.49 \pm 0.34$ mHz.

We note that since we are measuring an amplitude, and therefore a 
strictly positive quantity, this result is consistent with no 
sidereal variation at the 1-sigma level:  in the case where $\bar{\delta 
\nu}_{Z,\alpha}$ and $\bar{\delta 
\nu}_{Z,\beta}$ have zero mean value and the same variance $\sigma$, 
the probability distribution for $A$
takes the form $P(A) = A \sigma^{-2} \exp(-A^{2}/2\sigma^{2})$, which 
has the most probable value occurring at $A = \sigma$.

\section{Error analysis}
\label{sec.error}

In addition to our automated acquisition of Zeeman frequency data, we 
continuosly monitored the maser's external environment.
At every ten second step, in addition to applied frequency and maser 
beat period, we recorded room temperature, maser cabinet temperature, solenoid 
current, maser power, ambient magnetic field, and active Helmholtz coil 
current (see Sec. \ref{subsec.magnetics}).

\subsection{Magnetic systematics}
\label{subsec.magnetics}

\begin{figure}
\begin{center}
\includegraphics{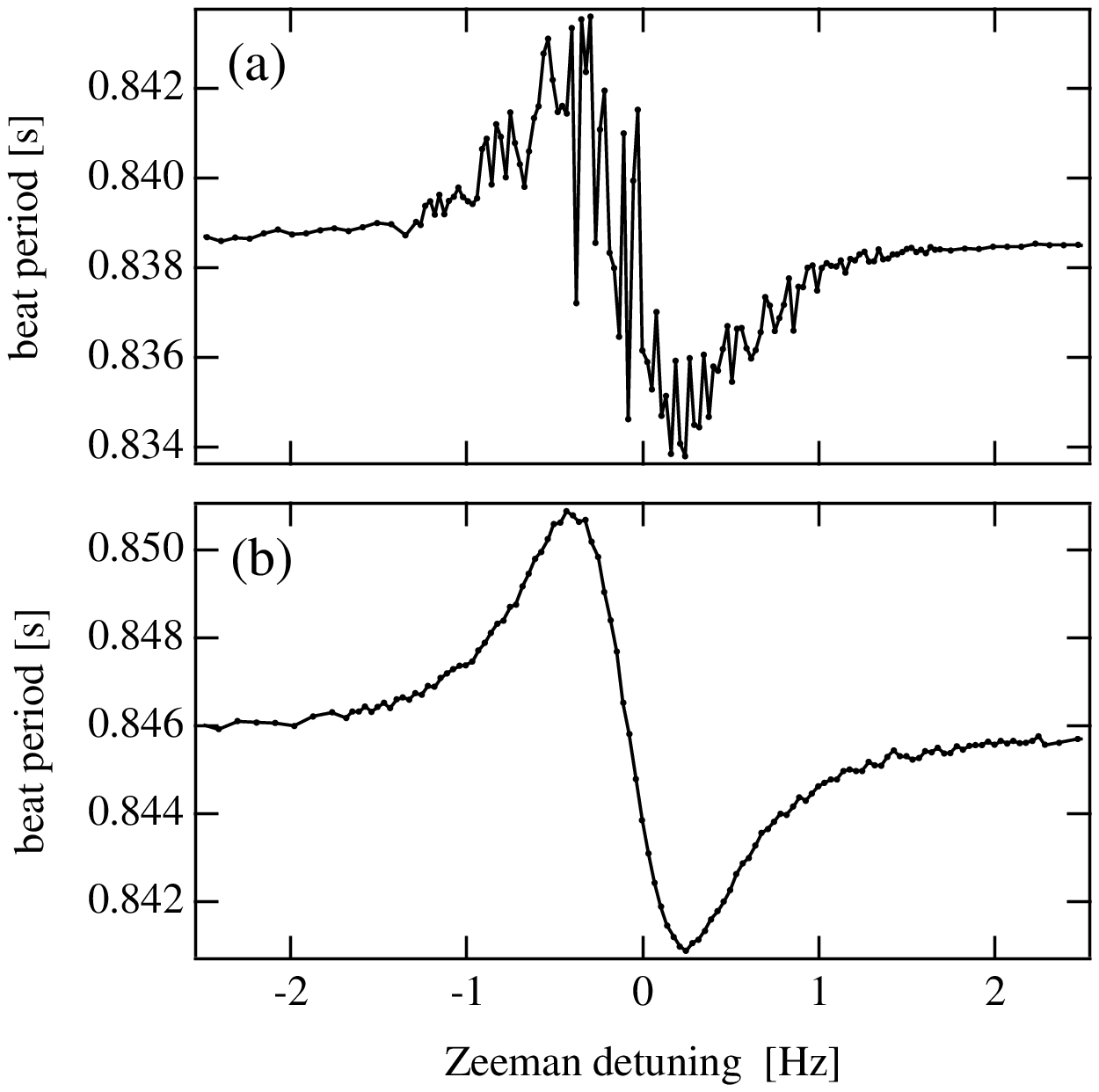}
\caption{(a) Zeeman scan without the active Helmholtz feedback loop.  
The noise on the data is due to the left and right shifting of the 
antisymmetric resonance as the Zeeman frequency shifts due to 3 mG 
ambient field fluctuations.  (b)  Zeeman scan with active Helmholtz 
control.  Ambient field fluctuations were reduced to less than 1 
$\mu$G.}
\label{fig.active.helm.data}
\end{center}
\end{figure}

The $F=1$, $m_{F}=\pm1$ Zeeman frequency depends to first-order on the 
z-component of the magnetic field in the storage bulb.  Thus, all 
external field fluctuations must be sufficiently screened to enable a 
sensitivity to shifts from Lorentz and CPT symmetry violation.  
The maser cavity and bulb are therefore surrounded by 
a set of four nested magnetic shields that reduce the ambient field by a 
factor of about 32,000.  We measure unshielded fluctuations in the 
ambient field 
of about 3 mG (peak-peak) during the day, and even when 
shielded, these add significant noise to a single Zeeman 
scan, as illustrated in Fig.~\ref{fig.active.helm.data}(a).  Furthermore, 
the amplitude of the field fluctuations is significantly reduced 
late at night, which could generate a diurnal systematic effect in our data.  

\begin{figure}
\begin{center}
\includegraphics{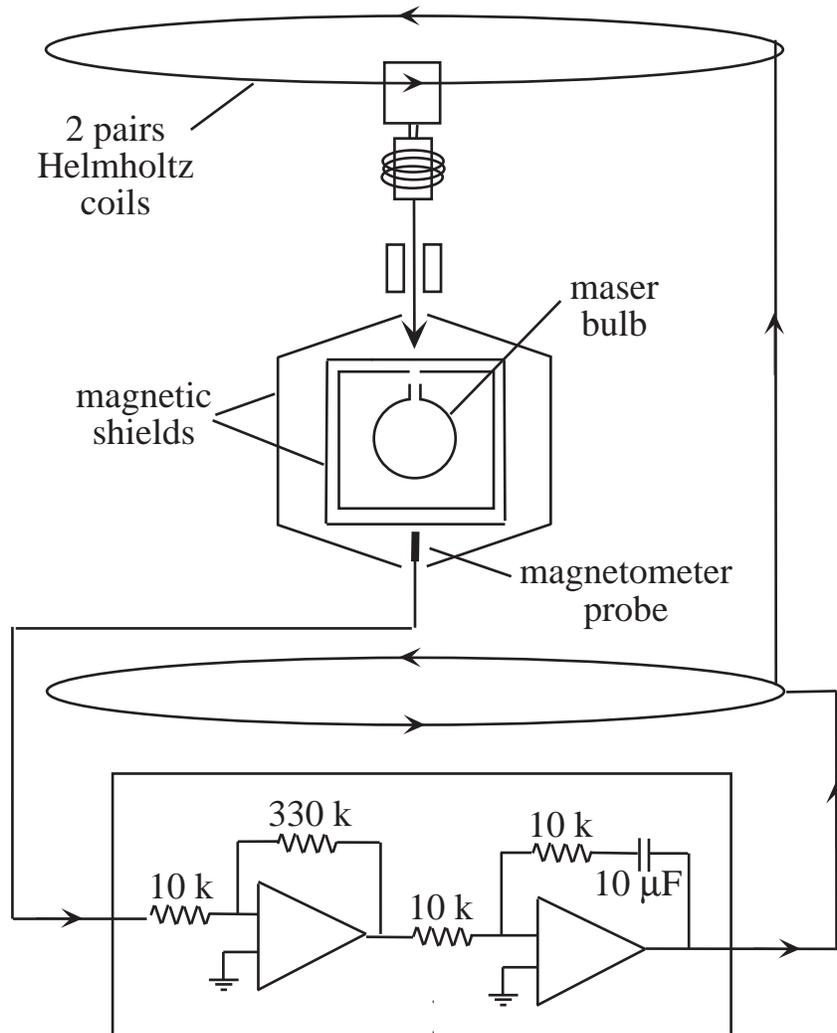}
\caption{Schematic of the active Helmholtz control loop.  
A large set of Helmholtz coils (50 turns) cancelled all but a residual 
$\sim$ 5 mG of the z-component of the ambient field. 
This residual field, detected with a fluxgate magnetometer probe, 
was actively cancelled by a servoloop and a second pair of Helmholtz coils 
(3 turns).  The servoloop consisted of a proportional stage 
(gain = 33), and integral stage (time constant = 0.1 s) and a 
derivative stage (time constant = 0.01, not shown).  
The overall time constant of the loop was about $\tau$ = 0.1 s.}
\label{fig.HelmholtzPID}
\end{center}
\end{figure}

To reduce the effect of fluctuations in the ambient magnetic field, we installed an 
active feedback system  (see Fig.~\ref{fig.HelmholtzPID}) consisting of two 
pairs of large Helmholtz coils (2.4 m diameter).  
The first pair of coils (50 turns) produced a uniform field that cancelled most 
of the z-component of the ambient field, leaving a residual field of around 5 mG.
A magnetometer probe that sensed the residual ambient field was placed 
partially inside the maser's magnetic shields 
near the maser cavity.  This probe had a sensitivity of $s$ = 1.7 mG/V.  
Due to its location partially inside the 
magnetic shields, the probe was screened by a factor of about six 
from external fields, reducing the sensitivity to $s^{\prime}$ = 0.3 mG/V, 
and producing a differential screening of 5300 
between the magnetometer probe and the atoms.  The magnetometer output was passed 
into a PID servo (Linear Research model LR-130), which contained a proportional 
stage (gain $G$ = 33), an integral stage (time constant T$_{i}$ = 0.1 
s) and a derivative stage (time constant = 0.01 s).  The correction 
voltage was applied to the second pair of Helmholtz coils (3 turns) 
which produced a uniform field ($p$ = 14 mG/V) along the z-axis to 
nullify the residual field and actively counter any field fluctuations.
Neglecting the small effect of the derivative stage, the overall time constant 
of this system was given by $\tau = T_{i} (1 + s^{\prime}/pG) \approx$ 0.1 
s, about 100 times shorter than the averaging time of our maser frequency 
shift measurements (10 s).

\begin{figure}
\begin{center}
\includegraphics{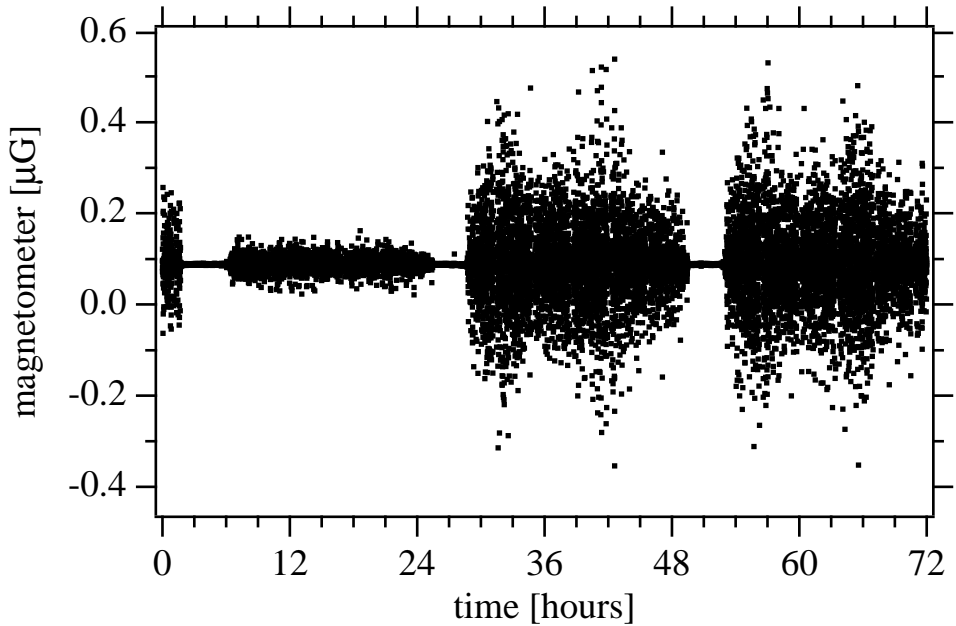}
\caption{Residual ambient magnetic field, after cancellation by the 
active Helmholtz control loop, sensed at the magnetometer 
probe.  Each point is a 10 s average.  These three days worth of data depict a Sunday, 
Monday and Tuesday, with the time origin corresponding to 00:00 Sunday.  
From these data it can be seen that for three hours every 
night the magnetic noise dies out dramatically, and that the noise 
level is significantly lower on weekends than weekdays.  
Nevertheless, with the active feedback system even the largest 
fluctuations (1 $\mu$G peak-peak) causes changes in the Zeeman frequency 
well below our sensitivity ($\Delta B = 1 \mu G \Rightarrow \Delta 
\nu_{Z} = 40 \mu Hz$).}
\label{fig.rfl.data}
\end{center}
\end{figure}

With this system we were able to further reduce ambient field
fluctuations at the magnetometer by a factor of 3,000.  The resulting
unshielded fluctuations were less than 1 $\mu$G peak-peak.  The
field recorded by the partially screened
magnetometer probe is shown in Fig.~\ref{fig.rfl.data}.  The noise on
a single Zeeman scan was reduced below our Zeeman frequency
resolution, as shown in Fig.~\ref{fig.active.helm.data}(b).  During
our Lorentz symmetry test, we monitored the field at the
magnetometer probe and placed a bound of $\sim$ 5 nG on
the sidereal component of the variation.  This corresponds to a shift of less than 0.2
$\mu$Hz on the hydrogen Zeeman frequency, three orders of
magnitude smaller than the sidereal Zeeman frequency bound measured.

The magnetometer \cite{fluxgate} used in the feedback loop was a fluxgate
magnetometer probe (RFL industries Model 101) which consisted of two 
parallel high-permeability magnetic cores each surrounded 
by an excitation coil (the excitation coils were wound in the opposite sense of 
each other).  A separate pickup coil was wound around
the pair of cores.  An AC current (about 2.5 kHz) in the excitation 
coils drove the cores into saturation, and, in the presence of any slowly 
varying external magnetic field oriented along the magnetic cores' axes, 
an EMF was generated in the pickup coil at the second and higher harmonics of 
the excitation frequency.  The magnitude of the time-averaged EMF was 
proportional to the external field.  The probe had a sensitivity of 
approximately 1 nG. 

Any Lorentz violating spin-orientation dependence of the energy of the electrons 
in the magnetic cores would induce a sidereal variation in the cores' 
magnetization and could generate, or mask, a sidereal variation in 
the hydrogen Zeeman frequency through the feedback circuit.  However, based on 
the latest bound on electron Lorentz violation \cite{adelberger1} (10$^{-29}$ 
GeV), the Lorentz violating shift would be less than
10$^{-11}$ G, far below the level of residual ambient field fluctuations.  
Also, the additional shielding factor 
of 5300 between the probe and the atoms
further reduced the effect of any Lorentz violating shift in the 
probe electrons' energies.

\begin{figure}
\begin{center}
\includegraphics{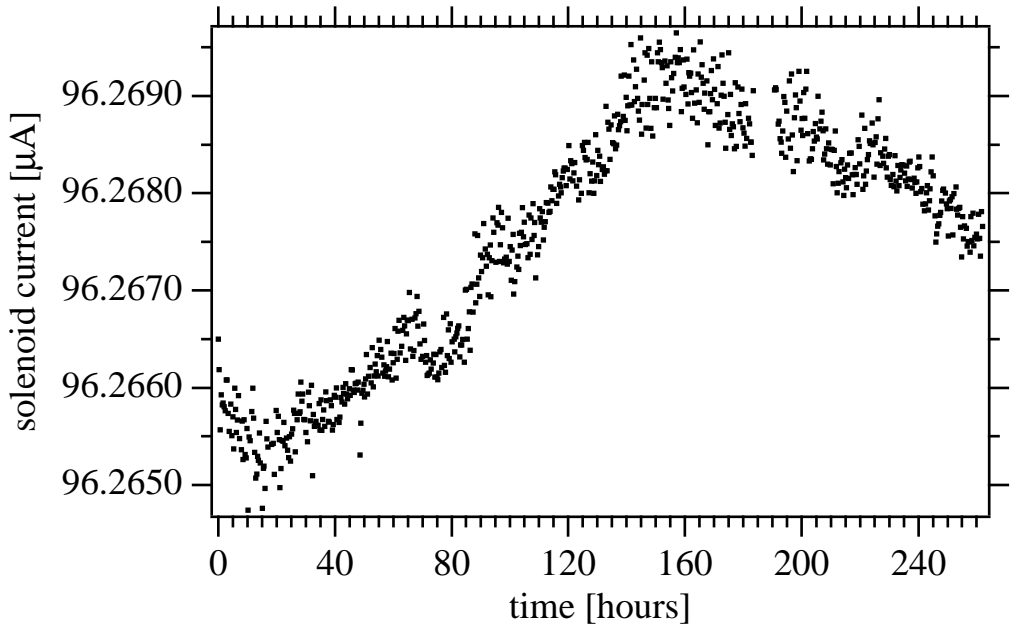}
\caption{Solenoid current during the first data run.  
  Each point is an average over one full Zeeman frequency measurement
  (18 mins).  Since the Zeeman frequency is directly proportional to
  the solenoid current, we subtracted these solenoid current drifts
  directly from the raw Zeeman data, using a measured calibration.  We
  find a sidereal component of 25 $\pm$ 10 pA to that correction,
  corresponding to a signal of 0.16 $\pm$ 0.08 mHz on the Zeeman
  frequency.  This systematic uncertainty has been included in our
  overall error analysis.}
\label{fig.solenoid.data}
\end{center}
\end{figure}

With the ambient field kept nearly constant near zero, the Zeeman
frequency was set by the magnetic field generated by the solenoid, and
hence by the solenoid current.  We monitored solenoid
current fluctuations by measuring the voltage across the
current-setting 5 k$\Omega$ resistor with a 5 1/2 digit multimeter
(Fluke model 8840A/AF).  By measuring the Zeeman frequency shift
caused by large current changes, we found a dependence of around 10
mHz/nA.  When acquiring Lorentz symmetry test data, we measured long
term drifts in the current of about 5 nA (see
Fig.~\ref{fig.solenoid.data}), significant enough to produce
detectable shifts in the Zeeman frequency.  Thus, we subtracted these
directly from the Zeeman data.  We measured a sidereal variation of 25
$\pm$ 10 pA on the solenoid current, corresponding to a sidereal
variation of 0.16 $\pm$ 0.08 mHz on the Zeeman frequency correction.
This systematic uncertainty in the Zeeman frequency was included in
the net error analysis, as described in
Sec.~\ref{subsec.error.analysis}.

\subsection{Other systematics}
\label{subsec.systems}

\begin{figure}
\begin{center}
\includegraphics{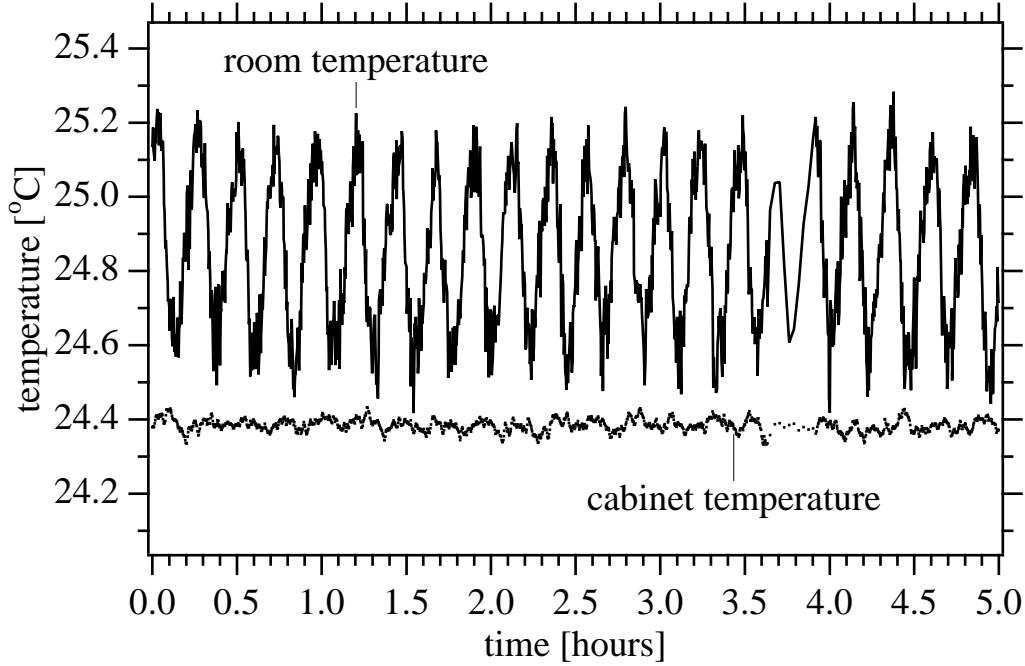}
\caption{Temperature data during the first run.  Each 
point is a 10 second average.  The top 
trace shows the characteristic 0.5\DegC \ peak-peak, 15 minute period 
oscillation of the room temperature.  The bottom trace shows the 
screened oscillations inside the maser cabinet.  The cabinet is 
insulated and temperature controlled with a blown air system.  In 
addition, the innermost regions of the maser, including the microwave 
cavity, are further insulated from the maser cabinet air temperature, 
and independently temperature controlled.  The residual temperature 
variation of the maser cabinet air had a sidereal variation of 0.5 mK, 
resulting in an additional systematic uncertainty of 0.1 mHz on the 
Zeeman frequency.  This value is included in the net error analysis.}
\label{fig.temp.data}
\end{center}
\end{figure}

The maser resided in a closed, temperature stabilized room where the 
temperature oscillated with a peak-peak amplitude of slightly 
less than 0.5\DegC \ with a period of around 15 minutes.  
The maser was contained in an insulated and thermally controlled 
cabinet, which provided a factor of five to ten shielding from the 
room, and reduced the fluctuations to less than 0.1\DegC \ 
peak-peak, as shown in Fig.~\ref{fig.temp.data}.  
By making large changes in the maser cabinet temperature 
and measuring the effect on the Zeeman frequency, we found a 
temperature coefficient of about 200 mHz/\DegC.  We believe this frequency 
shift was due mainly to the resistors which set the solenoid current, 
which had 100 ppm/\DegC\ temperature coefficients.  
We monitored the cabinet temperature and placed a bound 
on the sidereal component of the temperature fluctuations at 0.5 mK, which would 
produce a systematic sidereal variation of 100 $\mu$Hz on the Zeeman 
frequency, about a factor of 3 smaller than the measured limit on sidereal 
variation in Zeeman frequency.

\begin{figure}
\begin{center}
\includegraphics{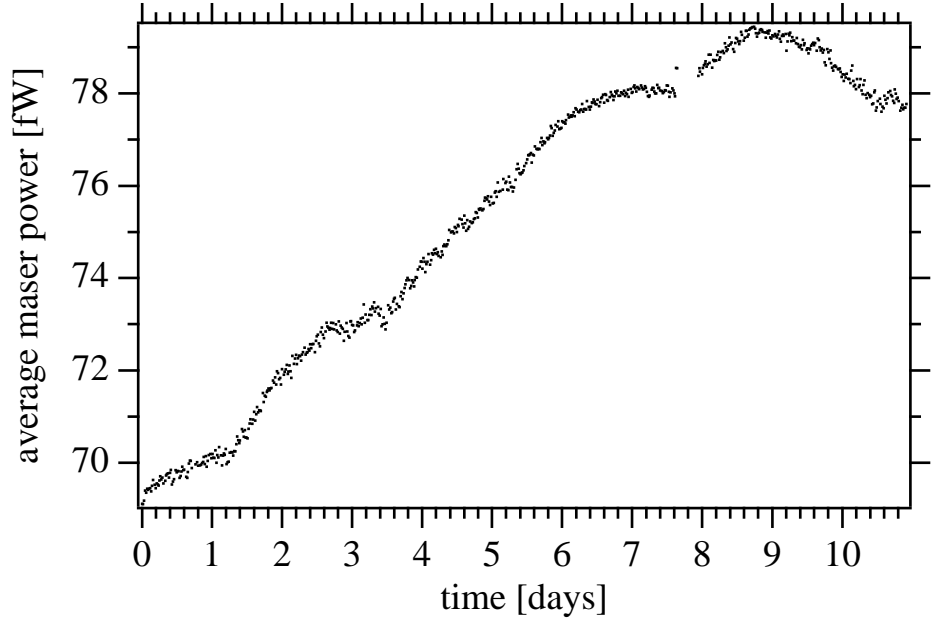}
\caption{Average maser power during the first data run.  
Each point is an average over one full Zeeman frequency measurement 
(18 mins).  We measure a sidereal variation in this power at 
less than 0.05 fW, leading to an additional systematic uncertainty in 
the Zeeman frequency of 0.04 mHz, which is included in the net 
error analysis.}
\label{fig.maspower.data}
\end{center}
\end{figure}

As mentioned in Sec. \ref{subsec.zeeman}, spin-exchange effects induce a small offset of the
Zeeman frequency given by Eqn.~\ref{eqn.analytic.andresen} from the
actual Zeeman frequency \cite{savard}.  This would imply that
fluctuations in the input atomic flux (and therefore the maser power)
could cause fluctuations in the Zeeman frequency measurement.  We
measured a limit on the shift of the Zeeman frequency due to large
changes in average maser power at less than 0.8 mHz/fW.  (Expected 
shifts from spin-exchange are ten times smaller than this level (Sec. 
\ref{subsec.zeeman}).  We believe the measured limit is related to 
heating of the maser as the flux is increased).  During
long-term operation, the average maser power drifted approximately 1 
fW/day (see Fig.~\ref{fig.maspower.data}).  The sidereal component of
the variations of the maser power were less than 0.05 fW, implying a
variation in the Zeeman frequency of less than 40 $\mu$Hz, an order of
magnitude smaller than our experimental bound for sidereal Zeeman
frequency variation.

\subsection{Final result}
\label{subsec.error.analysis}

We measured systematic errors in sidereal Zeeman frequency variation (as described in Secs.\ 
\ref{subsec.magnetics} and \ref{subsec.systems}) due to ambient magnetic field (0.2 $\mu$Hz), 
solenoid field (80 $\mu$Hz), maser cabinet temperature (100 
$\mu$Hz), and hydrogen density induced spin-exchange shifts (40 
$\mu$Hz).  Combining these errors in quadrature with the 0.34 mHz 
statistical uncertainty in Zeeman frequency variation, we find a sidereal variation 
of the $F=1$, $\Delta m_{F} = \pm 1$ 
Zeeman frequency in hydrogen of 0.44 $\pm$ 0.37 mHz at the 1-$\sigma$ 
level.  This 0.37 mHz bound corresponds to 1.5 
$\times$ 10$^{-27}$ GeV in energy units. 

\section{Discussion}
\label{sec.discussion}

\subsection{Transformation to fixed frame}
\label{subsec.latitude}

\begin{figure}
\begin{center}
\includegraphics{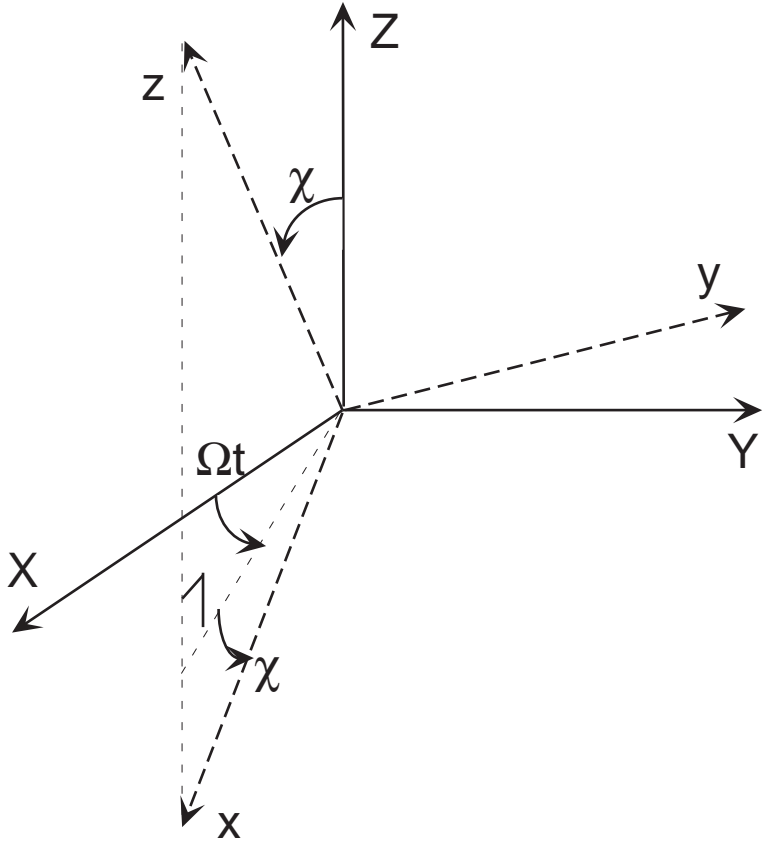}
\caption{Coordinate systems used.  The (X,Y,Z) set refers to a fixed 
reference frame, and the (x,y,z) set refers to the laboratory frame.  
The lab frame is tilted from the fixed Z-axis by our co-latitude, and it 
rotates about Z as the earth rotates.  The $\alpha$ and $\beta$ axes, 
described in Sec. \ref{sec.experiment}, 
span a plane parallel to the X-Y plane.}
\label{fig.fframe}
\end{center}
\end{figure}

Our experimental  bound of 0.37 mHz on sidereal variation of the 
hydrogen Zeeman frequency may be interpreted
in terms of Eqn.~\ref{eqn.Kost} as a bound on vector and tensor components 
of the standard model extension.  To make meaningful comparisons to other experiments, we
transform our result into a fixed reference frame.  Following the
construction in reference \cite{Kost.clock1}, we label the fixed frame
with coordinates (X,Y,Z) and the laboratory frame with coordinates
(x,y,z), as shown in Fig.~\ref{fig.fframe}.  We select the earth's
rotation axis as the fixed Z axis, (declination = 90 degrees). We then
define fixed X as declination = right ascension = 0 degrees, and fixed
Y as declination = 0 degrees, right ascension = 90 degrees.  With this
convention, the X and Y axes lie in the plane of the earth's equator.
Note that the $\alpha$, $\beta$ axes of Sec. \ref{subsec.ave}, also in the earth's equatorial
plane, are rotated about the earth's rotation axis from the X,Y axes
by an angle equivalent to the right ascension of 71$^{\circ}$ 7' longitude at 00:00 of November 19, 1999.

For our experiment, the quantization axis (which we denote z) was vertical in the lab frame,  
making an angle $\chi \approx$ 48 degrees 
relative to Z, accounted for by rotating the entire (x,y,z) system by $\chi$ 
about Y.  The lab frame (x,y,z) rotates about Z by an angle 
$\Omega t$, where $\Omega$ is the frequency of the earth's (sidereal) 
rotation.

These two coordinate systems are related through the transformation
  \begin{equation}
  \label{eqn.transform}
    \left( 
    \begin{array}{c}
        t \\ 
        x \\
        y \\
        z 
    \end{array}
    \right)
  = \left(
    \begin{array}{cccc}
        1 & 0 & 0 & 0 \\
        0 & \cos\chi \cos\Omega t & \cos\chi \sin\Omega t & -\sin\chi  \\
        0 & -\sin\Omega t & \cos\Omega t & 0 \\
        0 & \sin\chi \cos\Omega t & \sin\chi \sin\Omega t & \cos\chi
    \end{array}
    \right)
    \left( 
    \begin{array}{c}
        0 \\ 
        X \\
        Y \\
        Z 
     \end{array}
     \right) = \mathbf{T}
     \left( 
     \begin{array}{c}
        0 \\ 
        X \\
        Y \\
        Z 
     \end{array}
     \right).
  \end{equation}
Then, vectors transform as $\vec{b}_{lab} = {\mathbf T} \ \vec{b}_{fixed}$, 
while tensors transform as
${\mathbf d}_{lab} = {\mathbf T} \ {\mathbf d}_{fixed}  {\mathbf T}^{-1}$.

As shown in equation (\ref{eqn.Kost}), our signal depends on the following 
combination of terms (for both electron and proton):
\begin{equation}
    \tilde{b}_{3} = b_{3} - m d_{30} - H_{12}.
\end{equation}
Transforming these to the fixed frame, we see
\begin{eqnarray}
    b_{3} & = & b_{Z} \cos \chi + b_{X} \sin \chi \cos \Omega t + b_{Y} \sin 
    \chi \sin \Omega t, \nonumber \\
    d_{30} & = & d_{Z0} \cos \chi + d_{X0} \sin \chi \cos \Omega t + d_{Y0} \sin 
    \chi \sin \Omega t,  \\
    H_{12} & = & H_{XY} \cos \chi + H_{YZ} \sin \chi \cos \Omega t + H_{ZX} \sin 
    \chi \sin \Omega t, \nonumber
\end{eqnarray}
so our observable is given by
\begin{eqnarray}
    \label{eqn.btwiddle}
     \tilde{b}_{3} & =  &  \left( b_{Z} - m d_{Z0} - H_{XY} \right) \cos \chi \nonumber \\
                   & +  & \left( b_{Y} - m d_{Y0} - H_{ZX} \right) \sin \chi \sin \Omega t \\
                   & +  & \left( b_{X} - m d_{X0} - H_{YZ} \right) \sin \chi \cos \Omega t \nonumber.
\end{eqnarray}
The first term on the right is a constant offset, not bounded by our 
experiment.  The second and third terms each vary at the sidereal 
frequency.  Combining Eqn.~\ref{eqn.btwiddle} (for both e$^{-}$ and 
p) with Eqn.~\ref{eqn.Kost}, we see
\begin{eqnarray}
    |\Delta \nu_{Z}|^{2} & = & \left[ \left( b_{Y}^{e} - m_{e} d_{Y0}^{e} - H_{ZX}^{e} \right) 
                              + \left( b_{Y}^{p} - m_{p} d_{Y0}^{p} - H_{ZX}^{p} \right) \right]^{2} 
                                                        \frac{\sin^{2} \chi}{h^{2}} \\
                   & + & \left[ \left( b_{X}^{e} - m_{e} d_{X0}^{e} - H_{YZ}^{e} \right) 
                              + \left( b_{X}^{p} - m_{p} d_{X0}^{p} - H_{YZ}^{p} \right) \right]^{2} 
                                                        \frac{\sin^{2} 
                                                        \chi}{h^{2}} \nonumber.
\end{eqnarray}
Inserting $\chi$ = 48 degrees, we obtain the final result
\begin{equation}
    \sqrt{ \left( \tilde{b}_{X}^{e} + \tilde{b}_{X}^{p} \right)^{2} + 
    \left( \tilde{b}_{Y}^{e} + \tilde{b}_{Y}^{p} \right)^{2}  } = 
    \left( 3 \pm 2 \right) \times 10^{-27} \ GeV.
\end{equation}
Our 1-sigma bound on Lorentz and CPT violation of the proton and 
electron is therefore 2 $\times$ 10$^{-27}$ GeV.

\subsection{Comparison to previous experiments}
\label{subsec.compare}

We compare our result with other recent tests of Lorentz and CPT 
symmetry in Table \ref{tab.compare}.  Although our bounds are 
numerically similar to the 
those from the $^{199}$Hg/$^{133}$Cs experiment, the simplicity of the hydrogen atom 
allows us to place bounds directly on the electron and proton; 
uncertainties in nuclear structure models do not complicate the 
interpretation of our result.  The recent limit set 
by the torsion pendulum experiment of Adelberger et. al. 
\cite{adelberger3} on electron Lorentz and CPT violation casts 
our result as a clean bound on Lorentz and CPT violation of the proton.

\begin{table}
\begin{center}
\begin{tabular}{l c c c}
    Experiment & $\tilde{b}_{X,Y}^{e}$ [GHz] & $\tilde{b}_{X,Y}^{p}$ [GHz]  & 
$\tilde{b}_{X,Y}^{n}$  [GHz]\\ 
    \hline
    anomaly frequency of e$^-$ in Penning trap \cite{dehmelt1}   &  $10^{-25}$ & - & -  \\ 
    $^{199}$Hg and $^{133}$Cs precession frequencies 
    \cite{hunter.HgCs.lli} & $10^{-27}$ & $10^{-27}$ & $10^{-30}$ \\                                  
    \emph{this work} \cite{lli.Hexp} & $10^{-27}$ & $10^{-27}$ & - \\  
    spin polarized torsion pendulum \cite{adelberger3}  &  $10^{-29}$ & - & - \\
    dual species $^{129}$Xe/$^{3}$He maser \cite{bear} & - & - & $10^{-31}$ \\
\end{tabular}   
\caption{Electron, proton and neutron experimental bounds.}
\label{tab.compare}
\end{center}
\end{table}

\subsection{Future work}
\label{subsec.future}

To make a more sensitive measure of the sidereal variation 
of the Zeeman frequency in a hydrogen maser, it will be important 
to clearly identify and reduce the magnitude of the long term drifts of 
the Zeeman frequency.  Possible sources of these drifts are
magnetic fields near the maser bulb caused by stray currents in 
heaters or power supplies in the inner regions of the maser.  Also, 
the scatter of the Zeeman data points, believed to be due mainly to 
residual thermal fluctuations, should be reduced.
Both of these objectives could be accomplished by carefully rebuilding 
a hydrogen maser, with better engineered power and 
temperature control systems.
\section{Acknowledgments}

We gratefully acknowledge the encouragement of Alan Kosteleck\'{y}.  
Financial support was provided by NASA grant NAG8-1434 and ONR grant 
N00014-99-1-0501.  M.A.H. thanks NASA for fellowship support under the Graduate 
Student Researchers Program.


\bibliographystyle{prsty}

\begin{thebibliography}{10}
\bibitem{gnote}
Gauge invariance and renormalizability exclude the parameters $e_{\nu}$,
  $f_{\nu}$, and $g_{\lambda \mu \nu}$ in the standard model extension. We
  therefore neglected them relative to the other terms.
\bibitem{P8/P28}
The maser used for our Lorentz symmetry test, maser P-28 was only temporarily
  available for our use, as it was in our lab only to undergo several repairs.
  As a result, the amount of data we could acquire was limited. Maser P-8 is
  housed permanently in our laboratory, so much of our characterization was
  done with this maser. However, this maser was not suitable for a Lorentz
  symmetry test because it suffered from large, long term Zeeman frequency
  drifts, attributed to less magnetic shielding and larger extraneous fields
  (e.g., from heating elements).
\bibitem{maxchi}
Had we chosen the slope discontinuity with maximum reduced chi square, the
  total sidereal amplitude for this run would have been 1.1 mHz $\pm$ 0.4 mHz.
\end{thebibliography}

\end{document}